\documentclass[11pt]{article}

\usepackage{graphicx}
\usepackage{graphics}
\usepackage{amsmath}
\usepackage{amscd}
\usepackage{array}
\usepackage{amsfonts}
\usepackage{fancyhdr}
\usepackage{oldlfont}
\usepackage{latexsym,amssymb,amsmath}
\usepackage[T1]{fontenc}
\usepackage{multirow}
\usepackage{rotating}

\newtheorem{theorem}{Theorem.}

\newtheorem{corollary}[theorem]{Corollary.}

\newtheorem{definition}[theorem]{Definition.}

\newtheorem{lemma}[theorem]{Lemma.}

\newenvironment{proof}[1][Proof]{\textbf{#1.} }{\ \rule{0.5em}{0.5em}}

\def \Prod{\displaystyle\prod}

\def \be{\begin{eqnarray}}
\def \ee{\end{eqnarray}}
\def \b*{\begin{eqnarray*}}
\def \e*{\end{eqnarray*}}

\def \RR{{\mathbb R}}

\def \F{{{\mathcal F}}}

\def \xx{{\mathbf{x}}}

\def \[{[\,\!\![}
\def \]{]\,\!\!]}

\def \1{{\bf 1}}

\def\Ac{{\cal A}}

\def\Fc{{\cal F}}
\def\Gc{{\cal G}}
\def\Hc{{\cal H}}
\def\Ic{{\cal I}}

\def\bdelta{\vec{\delta}}
\def\bPhi{ \bar{\Phi}}

\begin{document}

\title{ {\LARGE
On break-even correlation: the way to price structured credit
derivatives by replication.}}
\medskip
\author{{\sc Jean-David Fermanian (CREST-ENSAE)}  \\ {\sc Olivier Vigneron (JP-Morgan)} \\}

\author{Jean-David Fermanian  \\\small CREST-ENSAE
                 \\\small jean-david.fermanian@ensae.fr
   \and Olivier Vigneron  \\\small JP-Morgan
    \\\small olivier.x.vigneron@jpmorgan.com
 }

\date{First version: April 2008. This version: \today}

\maketitle

\medskip
\begin{abstract}
We consider the pricing of European-style structured credit payoff
in a static framework, where the underlying default times are
independent given a common factor. A practical application would
consist of the pricing of nth-to-default baskets under the Gaussian
copula model (GCM). We provide necessary and sufficient conditions
so that the corresponding asset prices are martingales and introduce
the concept of ``break-even'' correlation matrix. When no sudden
jump-to-default events occur, we show that the perfect replication
of these payoffs under the GCM is obtained if and only if the
underlying single name credit spreads follow a particular family of
dynamics. We calculate the corresponding break-even correlations and
we exhibit a class of Merton-style models that are consistent with
this result. We explain why the GCM does not have a lot of
competitors among the class of one-period static models, except
perhaps the Clayton copula.
\medskip
\noindent

\medskip
\noindent


\end{abstract}

\newpage

\section{Introduction}

The risk management of structured credit products has become a key
issue for many financial institutions. Even if their payoffs are
only driven by the realization of default events, the value of these
products are exposed to credit spread volatility. The recent credit
crisis has highlighted the need for pricing models which can produce
dynamic hedging strategies that would replicate the payoff at
maturity of these products and, more importantly, reduce to zero the
``P\&L volatility'' of structured credit trading portfolios.

\medskip

The most common structured credit products are synthetic CDO
tranches (also known as CSO tranches). A standard modelling approach
to value CSO tranches is the one factor Gaussian copula model (GCM).
This model specifies directly the joint law of the underlying
default times (Li, 2000), where the dependence between these default
times is measured through a correlation parameter. With the marking
of correlation levels of base tranches $[0,K]$, $K\in [0,100\%]$,
GCM has become a market standard known as the ``base correlation''
model (see O'Kane and Livasey 2004, for instance).

\medskip

The Gaussian copula approach has been criticized for several
reasons. Firstly, the model does not define any credit spread or
default intensity dynamics. It only takes as a set of inputs current
credit spread curves, recoveries and a correlation. Secondly, the
common opinion is that CDO tranches are not priced through a
replication argument, contrary to the Black-Scholes model. D.
Li~\footnote{in {\em The Definitive Guide to CDOs}, G. Meisner (ed),
Risk Books, $p.71$} also wrote: ``The current copula framework gains
its popularity owing to its simplicity. However, there is little
theoretical justification of the current framework from financial
economics. We essentially have a credit portfolio model without
solid credit portfolio theory''. In our opinion, these criticisms
are partly due to a lack of understanding of the properties and
limitations of the GCM. In this article, we will consider the larger
class of static factor models, for which the default times are
assumed as independent conditionally on a random variable (the
so-called ``common factor''). We will explain to what extent and
under which conditions such copula models can be seen has pricing
models by replication, as in the standard Black-Scholes theory. In a
continuous spread model, we show that the payoff of an arbitrary
European basket credit derivative can be perfectly hedged under
particular spread dynamics and if the copula pricing model satisfies
some functional equation.

\medskip

The replication of CDO-type cash flows has generated a significant
amount of literature in the last years. Some authors have tried to
write explicit hedging strategies for some particular top-down
models, particularly when risks are driven mainly by sudden default
events (pure jump dynamics with contagion): Arnsdorf and Halperin
(2008), Frey and Backhaus (2008, 2009), Herbertsson (2008), Laurent
et al. (2008), among others. Then, the relevant hedging instruments
are most often instantaneous CDS (virtual instruments) and/or credit
indices. See the survey of Cousin, Jeanblanc and Laurent (2010).
Recently, Cousin and Jeanblanc (2010) have obtained general
theoretical results when the default events are ordered. None of
these authors deals with the case of static factor models,
particularly the market standard GCM. Others have studied
empirically the performances of various delta hedging strategies,
including the current practice: Cont and Kan (2008), Ammann and
Brommundt (2009), Cont et al. (2010), Cousin, Crepey and Kan (2010)
etc. In this article, we choose rather an opposite point of view: we
restrict ourselves on the diffusive part of the risk (spread risk),
but under standard market models and with individual CDS as hedging
instruments.

\medskip

In section~\ref{General}, we specify the framework and the
notations. We explain how the price given by a general factor model
can become a martingale. In section~\ref{GCMrevisited}, we
particularize the latter result in the case of the one-factor
Gaussian copula model (1FGCM). We exhibit the single credit spread
dynamics that are consistent with the pricing by replication of
credit basket structures in this model. We call ``break-even''
correlation the flat Gaussian copula correlation level to perfectly
replicate a given structured product with individual CDS, when no
sudden default occurs. Break-even correlations are written
explicitly as some functions of the spread volatilities and
correlations. Section~\ref{GCMrevisitedExtended} extends these
results in the general case of $p$-factor Gaussian copulas, and
introduces the concept of break-even correlation matrix. In
section~\ref{FPTD}, we calculate explicit hedging errors for all
First-$p$-To-Default securities. In section~\ref{StructuralModels},
we show the one-to-one correspondence between our spreads dynamics
and a class of structural models . Alternative factor models are
studied in section~\ref{AlternativeModels}. Finally,
section~\ref{Empiric} provides empirical illustrations of our
results, particularly break-even correlation levels, by simulating
spread trajectories and by using real market quotes.

\section{The framework}
\label{General}
First, let us describe our underlying market and technical assumptions.
To simplify, we will assume we are not exposed to any interest rate risk:

\medskip

Assumption (I): interest rates are zero now and in the future.

\medskip

The basic building blocks in the market are $n$ individual
$T$-maturity Credit Default Swaps, or, equivalently, $n$ survival
probabilities up to $T$. To be specific, for every name $i$ and
every maturity $T$, let the survival probabilities $Q_{it}(T)$ be
\begin{equation}
Q_{it}(T)=Q(\tau_i > T | {\mathcal F}_t):=\exp\left( - h_{it} .(T-t) \right).
\label{hazard}
\end{equation}
When there will be no ambiguity, such survival probabilities will be
denoted by $Q_{it}$ or even $Q_i$. In the equation above, $Q$
denotes a ``risk-neutral'' probability measure and  $h_{it}$ is the
hazard rate of $i$ between time $t$ and maturity $T$. From the
quantity $Q_{it}$, we can deduce the price of the $T$-maturity CDS
of name $i$: for every $i$ and $t$, $dCDS_{it} =
-dQ_{it}(T).(1-R_i),$ where $R_i$ is the constant recovery rate of
$i$. As a result, $Q_{it}$ can be considered as a traded asset and
our market will be defined as $(Q_{1t}(T),\ldots,Q_{nt}(T))_{t\in
[0,T]}$, endowed with its natural filtration $\F:=(\F_t)_{t\in
[0,T]}$, $\F_t=\sigma(Q_{iu}(T),u\leq t,i=1,\ldots,n)$. Note that
the filtration ${\mathcal F}$ contains all the relevant information
at time $t$, including past default events as well as past and
current credit spreads.

\medskip

Second, let's consider a credit basket product of $n$ underlying
names, $n\geq 2$. Its payoff depends only on the realizations of
default events in the underlying basket up to the maturity T of the
basket. We will further assume that all the default payments are
made at maturity and that the premium to enter this derivative
contract is paid upfront. Mathematically, we can write the payoff
using indicator functions:
$$ \psi(\delta_1(T),\ldots,\delta_n(T)), \;\;\;\; \delta_i(T) = \1(\tau_i \leq T), \; i=1,\ldots,n,$$
where $\tau_i$ is the default time of obligor $i$. For convenience,
set $\bdelta (t) = (\delta_1(t),\ldots,\delta_n(t))$. This framework
encompasses potentially all the European credit basket derivatives,
through an Arrow-Debreu type payoff decomposition.

\medskip

Placing ourselves within the framework of an arbitrage-free
financial market model, as in Bielecki and Rutkowski (2002) for
instance, the $t$-value of any attainable product is the expected
value of its future cash-flows under a risk-neutral measure $Q$.
Here, it will be
\begin{equation}
 V_t = \sum_{\bdelta(T)} \psi(\bdelta (T) ) Q(\bdelta(T)| {\mathcal F}_t).
\label{attainableV}
\end{equation}
In other words, $(V_t)_{t\in [0,T]}$ has to be a $\F$-martingale under $Q$.

\medskip

In order to derive tractable pricing and hedging of the basket
derivatives, we make a further assumption that will allow us to work
in the class of ``static'' factor models:

\medskip

Assumption (F1): At any time, conditionally on some random factor
$X\in \RR^p$ and the current market information, the underlying
default events $(\tau_i\leq T)$, $i=1,\ldots,n$, are independent.
The factor $X$ has got a density $f_X$ w.r.t. the Lebesgue measure
in $\RR^p$.

\medskip

With the latter assumption, we encompass most of the current market
models in the credit area (see for instance Burtschell et al. 2005).
Note that ``systemic'' factor $X$ depends on the current time $t$
implicitly, but we specify no dynamics concerning the successive
$X$. This factor $X$ could be multidimensional and when $X$ is a
vector, its values will be denoted in bold. When we restrict
ourselves to a one factor model, we will add this to our set of
assumptions:

\medskip

Assumption (F2): The dimension of systemic factor $X$ is one.

\medskip

Under (F1), the risk-neutral joint law of the default times is
defined by some intermediary quantities called ``conditional default
probabilities'' $p_{i|\xx}=Q(\tau_i \leq T | X=\xx, {\mathcal F}_t
)$, that depend on $Q_{it}$, $t$ and $\xx$ only. Similarly the
``conditional survival probabilities'' will be denoted by $q_{i|\xx}
:= 1-p_{i|\xx}$. It is proved in section~\ref{StructuralModels} that
$(p_{i|\xx})_{t\in [0,T]}$ is a martingale under $Q$ and for a
convenient extension of the filtration $\F$. There, $\xx$ will be
the proper $t$ realization of an adapted process in this extended
filtration.

\medskip

The $t$-value of the structured product is denoted by $V_t$. As usual in a factor model and under (F1),
\begin{equation}
 V_t = \sum_{\bdelta(T)} \psi(\bdelta (T) ) \int \Prod_{j=1}^n  p_{j|\xx}^{\delta_j}q_{j|\xx}^{1-\delta_j} f_X(\xx) \, d\xx ,
\label{FactorModelForm}
\end{equation}
where the summation is w.r.t. all the $n$-vectors of default
indicators. Note that $V_t= V(Q_{1t},\ldots,Q_{nt})$, and $V_t$ does not depend on time or
remaining maturity, except through the survival probabilities themselves.
Equivalently, there is no theta effect: $\partial V_t / \partial t =0$.

\medskip

The key issue we are interested in is the following one: Under which
conditions is the ``ad-hoc'' standard pricing
formula~(\ref{FactorModelForm}) a $\F$-martingale ? In other words,
can we write $ V_t= E_Q\left[ \psi(\bdelta) | {\mathcal F}_t
\right]$, as we can for any attainable claim
(see~(\ref{attainableV}))?

\medskip

Note that the pricing formula in equation~(\ref{FactorModelForm})
does not depend on the volatilities of the default intensities
and/or probabilities, due to the static nature of the underlying
model~\footnote{This is different from the Black-Scholes model,
where the pricing of options involves some volatility parameters.
Potentially, the latter ones are different from the ``true''
realized volatilities, inducing some hedging errors: see Carr and
Madan (2001). In our case, such a discrepancy between the risk
neutral parameters and the realized ones will appear below, but
through correlation levels : see section~\ref{GCMrevisited}.}. It
does not mean that the volatilities $\sigma_{it}$ do not matter to
calculate the price $V_t$. Indeed, we will see that they influence
the choice of the ``right'' pricing parameters, particularly the
pricing correlation in the GCM.

\medskip

Assumption (A): no default event will occur in the time interval
$[0,T)$.

\medskip

Apparently, the latter assumption is a bit provocative: it assumes
we are living in a credit world without credit events except at the
maturity $T$. However, this does not preclude that the spread of any
name becomes so wide that it trades close to default within that
period. Also, for short/medium term horizons and for investment
grade portfolios (like Master indices), the assumption (A) will be
acceptable and is equivalent to consider that the structured
product, while in a trading book, is exposed to spread risks mainly.
It could be stressed that, historically, most of the default events
have been anticipated by the main market dealers some weeks before
their official announcements with the price of the CDS trading near
the recovery value. Additionally, we assume that the survival
probabilities $Q_{it}(T)$ are Ito processes. Thus, we will live in a
Black-Scholes world.

\medskip

Assumption (Q): Under the historical measure $P$, the survival probability process of name $i$ up to maturity $T$ is given by
$dQ_{it}(T) = \mu_{i,t,T}dt + \sigma_{i,t,T} dW_{it},$
when $i$ is not defaulted, and $Q_{it}(T)=0$ else. The volatility and drift processes above are ${\mathcal F}$-adapted.
Obviously, the processes $(W_{it})$ are correlated $\Fc$-Brownian motions under $P$: $E_P[dW_{it}.dW_{jt}]=\rho_{ij}dt$, $i\neq j$.
We will call $\rho_{ij}$ the ``spread correlation'' between the names $i$ and $j$.

\medskip

As deduced from standard arguments, the risk neutral $Q$ can be
assumed to be the risk-neutral measure given by the Girsanov's
theorem. Therefore, under $Q$, the survival probabilities follow the
processes $dQ_{it}(T) =\sigma_{i,t,T} dW_{it},$ when $i$ is not
defaulted, and $Q_{it}(T)=0$ else. Here, the processes $(W_{it})$
are correlated $\Fc$-Brownian motions under $Q$, and
$E_Q[dW_{it}.dW_{jt}]=\rho_{ij}dt$, $i\neq j$. Since there will be
no ambiguity, $\sigma_{i,t,T}$ is denoted by $\sigma_{it}$ simply.
We denote $\Sigma^S :=[\rho_{ij}]_{1\leq i,j \leq n}$, where
$\rho_{ii}=1$. We prove

\begin{theorem}
\label{thdpit2} Under the assumptions (I) (Q) (A) (F1) (F2),
\begin{align*}
dV_t &= (\ldots)\cdot d\vec{W}_t +
\frac{dt}{2}\sum_{\bdelta(T)} \psi(\bdelta (T) ) \sum_{(i,j),i<j} \int \Prod_{k\neq i,j}
 p_{k|x}^{\delta_k}q_{k|x}^{1-\delta_k}.(2 \delta_i-1)(2 \delta_j-1) \\
 &\cdot \left[ 2 \sigma_{it} \sigma_{jt} \rho_{ij}
 \frac{\partial p_{i|x}}{\partial Q_i}\cdot \frac{\partial p_{j|x}}{\partial Q_j} f_X(x)
 -\sigma_{it}^2  \frac{\partial p_{j|x}}{\partial x} \chi_i(x) -
 \sigma_{jt}^2 \frac{\partial p_{i|x}}{\partial x} \chi_j(x) \right]  \, dx +\eta
 \end{align*}
where
$$ \chi_i(x):=\int_{-\infty}^x
\frac{\partial^2 p_{i|u}}{\partial^2 Q_i}  f_X(u) \, du,$$
and
$$\eta := \frac{dt}{2} \sum_{\bdelta(T)} \psi(\bdelta (T)) \sum_{i=1}^n (2 \delta_i-1)
 \frac{\sigma_i^2 }{2}
\left[ \Prod_{j,j\neq i}  p_{j|x}^{\delta_j}q_{j|x}^{1-\delta_j} \chi_i(x)  \right]_{-\infty}^{+\infty}.$$
\end{theorem}
When there are two names only in the basket, the product over $k$ above is just replaced by the constant one.
See the proof in the appendix~\ref{thgeneral}.
The last term residual term $\eta$ is zero
in a lot of models, particularly the Gaussian copula one. Actually, $\eta=0$ when $\lim_{|x|\rightarrow +\infty} \chi_i(x) = 0$
for every name $i$.

\medskip

Therefore, under the ``no default'' assumption (A), it is possible
to get a strictly flat P\&L by delta hedging any structured product
with $T$-maturity individual CDS if and only if the following
partial differential equation is satisfied: for every couple of
indices $(i,j)$, $i\neq j$, for every time $t$ and every $x$,
\begin{equation}
2\sigma_{it} \sigma_{jt} \rho_{ij}
\frac{\partial p_{i|x}}{\partial Q_i}\cdot \frac{\partial p_{j|x}}{\partial Q_j} f_X(x)
=\sigma_{it}^2  \frac{\partial p_{j|x}}{\partial x} \chi_i(x) +
 \sigma_{jt}^2  \frac{\partial p_{i|x}}{\partial x} \chi_j(x).
\label{GeneralCondition}
\end{equation}

Note that, in general, traders try to be hedged simultaneously
against spread moves and against jump-to-defaults. This means they
have to trade two types of hedging instruments, typically two CDS
per name with different maturities. In that case, it is no more
possible to get a flat P\&L, because the trader will pay carry to be
protected against default events. In average and in the long-term,
this additional cost is zero, but not between two successive times.

\section{The Gaussian Copula revisited}
\label{GCMrevisited} It is now fruitful to consider the Gaussian
copula model under the framework of section~\ref{General}. Indeed,
it is still the market standard, even if it has been criticized,
particularly during the credit crisis 2007-2009.

\medskip

Assumption (Cop): To price our European structured product, the
framework is the one-factor Gaussian copula model. The correlation
between the $i$-th asset value and the systemic factor $X$ is
denoted by $\rho_i$.

\medskip

Since we consider payments at maturity only, the timing of default events and term structures of credit spreads do not matter to price the
structured product. As a consequence, (Cop) is equivalent to define the joint law of the default events $\vec{\delta}(T)$ only.
In general, both approaches are not equivalent. Typically, the so-called ``base correlation'' methodology specifies the joint law of default
events at several maturities, but it is not possible to infer from them the joint law of default times, except when the correlation
levels are the same for all maturities.

\medskip

As a consequence of (Cop), the correlation between two different asset values $X_i$ and $X_j$ will be $\rho_i\rho_j$.
Set $ \bPhi = 1- \Phi$ and
$$d_j := d_j(T,x) := \frac{ \bPhi^{-1} (Q_{jt}) - \rho_j x}{\sqrt{1-\rho_j^2}}\cdot$$
Under (Cop), the ``conditional default probability'' of name $j$ by time $T$ is by definition
\begin{equation}
 p_{j|x} = \Phi\left( \frac{\bPhi^{-1}(Q_{jt})- \rho_j x}{\sqrt{1-\rho_j^2}} \right)= \Phi(d_j),
\label{GCMSpecif}
\end{equation}
and $q_{j|x} = 1- p_{j|x} =\bPhi(d_j).$

\medskip

It is proved in the appendix~\ref{technical} that the function $\chi_j$ of theorem~\ref{thdpit2} is now
\begin{equation}
 \chi_j(x) = \frac{\rho_j \phi(x)}{\phi(\Phi^{-1}(Q_{jt}))}\cdot \frac{\partial p_{j|x}}{\partial Q_j}\cdot
\label{technicalCopGauss}
\end{equation}
Equation~(\ref{technicalCopGauss}) is a noteworthy and a nice
property of the Gaussian copula specification. Note that $\chi_j(x)$
tends to zero when $|x|$ tends to the infinity. We deduce that the
``noisy'' term $\eta$ in theorem~\ref{thdpit2} is zero.

\begin{theorem}
\label{theocopG} Under the assumptions (Cop) (I) (A) (Q) (F1),
\begin{align*}
dV_t &=(\cdots)\cdot d\vec{W}_t+
\frac{dt}{2}\sum_{\bdelta(T)} \psi(\bdelta (T) ) \sum_{i<j} \int \Prod_{k\neq i,j}
 p_{k|x}^{\delta_k}q_{k|x}^{1-\delta_k}.(2 \delta_i-1)(2 \delta_j-1) \\
 &\cdot \left[ 2 \beta_{it} \beta_{jt} \rho_{ij}
 -\rho_i \rho_j \beta_{it}^2  - \rho_i \rho_j \beta_{jt}^2  \right]
 \frac{\phi(d_i)\phi(d_j) \phi(x)}{\sqrt{1-\rho_j^2}\sqrt{1-\rho_i^2}}\, dx ,
\end{align*}
where
$$ \beta_{it} := \frac{\sigma_{it}}{\phi(\Phi^{-1}(Q_{it})) }\cdot $$
\end{theorem}

In practice, it is highly valuable to exhibit the ``right'' values of pricing parameters, i.e. the values under which $(V_t)$ becomes a $\F$-martingale under $Q$. Then, without any default default before $T$, we can hedge our structured product dynamically with $T$-maturity CDSs'.
Under (Cop), the single parameters are the ``beta'' factors
$\rho_i$, $i=1,\ldots,n$.
Remind that we know the ``real-world'' correlation
parameters $\rho_{ij}$. This leads us to introduce the concept of break-even correlation.

\begin{definition}
A given structured credit security is priced under the one-factor
Gaussian copula framework with a constant parameter $\rho$. This
security is hedged dynamically with individual $T$-maturity CDS and
no default occurs during the time interval we consider. Then, the
value of the copula (flat) correlation that induces a zero P\&L in
this time interval and for a realized trajectory of individual
credit spreads is called a ``break-even'' correlation. When the time
interval is infinitesimal, we will be calling it ``instantaneous
break-even'' correlation. When its value is independent of the
(random) credit spread trajectories, it will be called a ``universal
break-even'' correlation.
\end{definition}

For us, the term ``break-even'' correlation is related to $\rho_{BE,t}^2$ and not to $\rho_{BE,t}$. The latter quantity will be called ``beta factor'' rather.
This concept has been introduced first in~\cite{StratReport} for First-To-Default. A break-even correlation is comparable to a
base correlation. Its squared root is the so-called ``beta''-factor used by practitioners, i.e. the correlation between an individual asset value and the systemic random factor in the one-factor Gaussian copula.

\medskip

Instead of looking for a single break-even correlation number and
flat correlation structures, we can extend these definitions
straightforwardly to deal with one-factor pricing structures. Then,
the beta factors $\rho_i$, $i=1,\ldots,n$, that induce a zero P\&L
in a given time interval and for a realized trajectory of individual
credit spreads will be called ``break-even beta factors''.

\medskip

Invoking theorem~\ref{theocopG}, the break-even correlation
$\rho_{BE,t}^2$ in the infinitesimal time interval $[t,t+dt]$ has to
be a linear combination of the pairwise spread correlations:
$$  \rho_{BE,t}^2 = \sum_{(i,j),i\neq j} w_{ij,t} \rho_{ij} ,$$
where the weights $w_{ij,t}$ involve individual survival
probabilities at $t$, instantaneous volatilities $\sigma_{it}$ and
the payoff functional~\footnote{Actually, the break-even correlation
itself is hidden in $w_{ij}$ through $d_i$ and $d_j$. This means we
have to solve an implicit function to calculate break-even
correlations.}. The weights $w_{ij,t}$ could be negative because of
the default indicator functions $\delta_k$. Thus, the existence of
$\rho_{BE,t}$ is not guaranteed. The only way to be sure that there
exist a break-even correlation, or at least break-even beta factors
is to cancel all the terms
$$ 2 \beta_{it} \beta_{jt} \rho_{ij}
 -\rho_i \rho_j \beta_{it}^2  - \rho_i \rho_j \beta_{jt}^2, \;\; i\neq j, $$
for every $\beta_{it}$ and $\beta_{jt}$.
If it can be done, we find universal break-even beta factors. And they will be the same whatever the payoff $\psi(\cdot)$, because the $\beta_{it}$ terms depend on the spread dynamics only.
This intuition will be formalized in the next theorem. Before stating this result, we need to define the names that have really an influence on the realized payoffs of the
structured product.

\medskip

Assumption (P): for every name $i$ in the basket, there exists another index $j\neq i$ and some indicators $\delta^*_k$, $k\neq i,j,$ such that
\begin{equation}
 \psi(1,1,\bdelta^*_{-(i,j)}) - \psi(1,0,\bdelta^*_{-(i,j)}) - \psi(0,1,\bdelta^*_{-(i,j)})
+ \psi(0,0,\bdelta^*_{-(i,j)}) \neq 0.
\label{payoffrelevant}
\end{equation}
In the equation above, the first (resp. second) index is related to name $i$ (resp. $j$), and $\bdelta_{-(i,j)}$ is the $n-2$-vector of
indicators $\delta_k$, $k\neq i,j$.

\medskip

The equation~(\ref{payoffrelevant}) means: the knowledge of $i$'s
default is informative to evaluate the change of the final payoff
induced by $j$'s behavior, under some scenario of other default
events. Such a technical condition is satisfied in practice by all
the usual credit derivatives, particularly First-To-Default
securities. Since every structured credit product can be seen as a
linear combination of First-To-Default on some sub-portfolios (see
Brasch 2006), (P) is satisfied generally when $i$ and $j$ take
really part of the payoff $\psi$.

\medskip

Moreover, to state the necessity part in the next theorem, we need another technical condition.

\medskip

Assumption (B): for every name $i$ in the basket, the function
$Q_i\mapsto \sigma_{it}(Q_i)/\phi(\Phi^{-1}(Q_i))$ from $(0,1)$ to $\RR^+$ is bounded from above.

\medskip

The latter assumption avoids some pathological behaviors, when
survival probabilities become close to zero or one. It provides a
type of upper bound in terms of volatility explosion in the tails.

\medskip

Now, we can state a striking result: we are able to exhibit the
dynamics of $Q_{it}$ (under the historical and risk-neutral
measures), that allow perfect replication in our framework. In
section~\ref{UniqueDynamics}, we prove the following result.

\begin{theorem}
\label{Unicity}
Under the assumptions (Cop) (I) (Q) (F1) (A) (P) (B), any credit basket derivative can be perfectly hedged against credit spread variations if and only if
\begin{enumerate}
\item[(i)] the dynamics of the survival probabilities under $P$ are given by
\begin{equation}
dQ_{it} =  \bar\sigma_{i}\xi_t \phi(\Phi^{-1}(Q_{it})) dW_{it} +\mu_{it} dt,
\label{BEDynamics}
\end{equation}
for every index $i$. Above, $ \bar\sigma_{i}$ denotes some positive constant and $\xi$ a deterministic function such that $\lim_{t\rightarrow T}
\int_0^t \xi_{u}^2 \, du = +\infty$.
\item[(ii)] there exist some real numbers $\rho_i^S$ such that
\begin{equation}
 \rho_{ij} = E_P[ dW_{it}.dW_{jt}]/dt  = \left[\frac{\bar\sigma_{i}^2 + \bar\sigma_{j}^2}{2\bar\sigma_{i}\bar\sigma_{j}}\right]\rho_i^S\rho_j^S, \;\; \forall i\neq j.
\label{CondExistenceBEC}
\end{equation}
\item[(iii)] for every index $i$, the correlations that are
used for the pricing of the structured product, in the one-factor Gaussian copula model, are given by the relations $ \rho_{i} = \rho_i^{S}.$
\end{enumerate}
\end{theorem}

\begin{corollary}
Under the assumptions of theorem~\ref{Unicity}, assume that the matrix of spread correlations $\Sigma^S$ is one-factor, i.e.
there exist coefficients $\rho_i^S$ such that $\rho_{ij}= \rho_i^S\rho_j^S$ for all $i\neq j$.
Then, if the coefficients $\bar\sigma_{it}$ are the same for every index $i$, the pricing by replication under the 1FGCM is possible.
In this case, $\rho_i = \rho_i^S$.
\end{corollary}

In general, an arbitrary correlation matrix $\Sigma^S$ (even
one-factor) and arbitrary deterministic ``volatility'' coefficients
$\bar\sigma_{i}$ do not imply the existence of ``break-even''
pricing coefficients $\rho_i$. Indeed, we are not sure that there
exist coefficients $\rho_i^S$ that satisfy the key
equation~(\ref{CondExistenceBEC}). In other words, the correlation
matrix $\Sigma^S$ must have got a particular structure, given
by~(\ref{CondExistenceBEC}), so that there exist break-even beta
factors. This family of correlation matrices is parameterized by the
``volatility'' coefficients $\bar\sigma_{i}$, except when all of
them are equal.

\medskip

Actually, consider the matrix $\tilde{\Sigma}^S$, obtained by reverting equation~(\ref{CondExistenceBEC}), i.e.
$$ \tilde{\Sigma}^S := \left[ \rho_{ij}\frac{2\bar\sigma_i\bar\sigma_j}{\bar\sigma_i^2 + \bar\sigma_j^2}  \right].$$
Fortunately, $\tilde{\Sigma}^S$ is a correlation matrix.
This is proved incidentally in section~\ref{StructuralModels}.
Nonetheless, we are not insured that this correlation matrix is
one-factor. Thus, the existence of the coefficients $\rho_i^S$ is
not guaranteed.

\medskip

In section~\ref{StructuralModels}, we will see that, typically, the
classical one-period structural model induces the choice $\xi_{t} =
(T-t)^{-1/2}$ and $\bar\sigma_i =1$. In this case, the only
discrepancy between the underlying spread dynamics is due to
different survival probabilities $Q_{it}$ and different spread
correlations. Now, we have stated that perfect replication may be
compatible with an additional level of heterogeneity between spread
volatilities, i.e. when the $\bar\sigma_{i}$ are different among the
underlying names. For instance, all other things being equal, when a
``volatility'' factor $\bar\sigma_i$ goes up (resp. down), then the
``break-even'' beta factor of $i$ goes down (resp. up), so that
equation~(\ref{CondExistenceBEC}) remains satisfied.

\medskip

Nicely, equation~(\ref{BEDynamics}) can be solved explicitly.
Indeed, it can be checked that the solution of the latter equation is

\begin{align}
Q_{it} &= \Phi\left(\exp(\frac{\bar\sigma_{i}^2 \int_0^t \xi_u^2 \, du}{2})
 \Phi^{-1}(Q_{i0}) +  \int_0^t \exp(\frac{\bar\sigma_{i}^2 \int_u^t \xi_s^2  \, ds}{2})\{ \bar\sigma_{i} \xi_u \, dW_{iu}  \right. \nonumber \\
& \left. + \frac{\mu_{iu}}{\phi(\Phi^{-1}(Q_{u}))} \, du\} \right),
\label{SolQ}
\end{align}
where $(W_{it})$ is a Brownian motion under $P$. Note that the
consistency condition $\lim_{t\rightarrow T} Q_{it}(T)  \in \{0,1\}$
is satisfied through the property $\lim_{t\rightarrow T} \int_0^t
\xi_u^2 \, du = +\infty$. Finally, note that the previous individual
dynamics can be rewritten easily in terms of hazard rates (or credit
spreads, approximately) $h_{it}$, defined by $
Q_{it}:=\exp(-h_{it}.(T-t))$.

\section{Extension to the $p$-factor Gaussian copula model}
\label{GCMrevisitedExtended}

Now, we extend the results of the previous section to deal with more general Gaussian copulas, i.e. not only one-factor.
By definition, this means that the default events are
independent w.r.t. a $p$-dimensional standard Gaussian random variable $X$, and
that there exist vectors $\rho_i\in \RR^p$, $i=1,\ldots,n$, such that ``conditional default probabilities'' are
\begin{equation}
p_{i|\xx} = Q(\tau_i \leq T |X=\xx, \Fc_t) = \Phi\left( \frac{\bPhi^{-1}(Q_{it})- \rho_i' \xx}{\sqrt{1-|\rho_i|^2}} \right):= \Phi(d_i),
\label{GenGCMSpecif}
\end{equation}
for all $i=1,\ldots,n$ and $\xx\in\RR^p$, and with our previous
notations. In the equation above, $|\rho_i|$ denotes the Euclidian
norm of the $p$-vector $|\rho_i|$. Obviously, joint default
probabilities are obtained by an integration w.r.t. the factor density
$f_X$, where $f_X:\RR^p \rightarrow \RR^+$, $\xx \mapsto
\prod_{k=1}^p\phi(x_k)$. In this framework, the pricing parameters
are the vectors $\rho_i\in \RR^p$, $i=1,\ldots,n$. They allow to
build a correlation matrix $\Sigma$ of rank $p$, for pricing
purpose. It is defined by
$\Sigma_{i,j} = \rho'_i \rho_j$, $i\neq j$ and $\Sigma_{ii}=1$.
Thus, in this section, we assume

\medskip

Assumption (pCop): To price any European structured product, the framework is the $p$-factor Gaussian copula model,
for some $p \in \{1,\ldots,n\}$.

\medskip

When $p\geq 2$, we cannot invoke theorem~\ref{thdpit2} anymore. Indeed, it is based
on a single integration by parts argument. When $x$ is replaced by a vector $\xx$, it is far
from obvious to exhibit the relevant univariate variable of integration, and then to find the new functions $\chi_i$.
Nonetheless and nicely, we prove here that the results of theorems~\ref{theocopG} and~\ref{Unicity} are still available.

\begin{theorem}
\label{GentheocopG}
Under the assumptions (pCop) (I) (A) (Q) (F1),
\begin{align*}
dV_t &=(\cdots)\cdot d\vec{W}_t+
\frac{dt}{2}\sum_{\bdelta(T)} \psi(\bdelta (T) ) \sum_{i<j} \int \Prod_{k\neq i,j}
 p_{k|\xx}^{\delta_k}q_{k|\xx}^{1-\delta_k}.(2 \delta_i-1)(2 \delta_j-1) \\
 &\cdot \left[ 2 \beta_{it} \beta_{jt} \rho_{ij}
 -\rho_i' \rho_j \beta_{it}^2  - \rho_i'\rho_j \beta_{jt}^2  \right]
 \frac{\phi(d_i)\phi(d_j)\phi(\xx)}{\sqrt{1-|\rho_j|^2}\sqrt{1-|\rho_i|^2}}\prod_{l=1}^p \phi(x_l)\, dx_l ,
\end{align*}
where
$$ \beta_{it} := \frac{\sigma_{it}}{\phi(\Phi^{-1}(Q_{it})) }\cdot $$
\end{theorem}

See the proof in appendix~\ref{Proof_GentheocopG}. Assume we know
the ``real-world'' correlation parameters $\rho_{ij}$ and the
volatility parameters $\beta_{it}$. This leads to the concept of
break-even correlation matrix.

\begin{definition}
A given structured credit security is priced under the $p$-factor Gaussian copula framework. This security is hedged dynamically with individual $T$-maturity
CDS and no default occurs during the time interval we consider.
Then, the $p$-factor correlation matrix that is used for pricing purpose and that
induces a flat P\&L in this time interval and for a realized trajectory of individual credit spreads
 is called a ``break-even'' correlation matrix.
When its value is independent of the (random) credit spread
trajectories, it will be called a ``universal break-even''
correlation matrix.
\end{definition}

Note that this matrix does not always exist for arbitrary spread correlations $\rho_{ij}$ and arbitrary volatilities $\sigma_{it}$.
As in the univariate case, we can detail necessary and sufficient conditions for the existence of a break-even correlation matrix.
With the same reasoning as in theorem~\ref{Unicity}, we find the single dynamics that are compatible
with perfect replication of credit structured products.
\begin{theorem}
\label{GenUnicity}
Under the assumptions (pCop) (I) (Q) (F1) (A) (P) (B), consider a particular credit basket derivative.
It can be perfectly hedged against credit spread variations if and only if
\begin{enumerate}
\item[(i)] the dynamics of survival probabilities under $P$ are given by
\begin{equation}
dQ_{it} =  \bar\sigma_{i}\xi_t \phi(\Phi^{-1}(Q_{it})) dW_{it} +\mu_{it} dt,
\label{GenBEDynamics}
\end{equation}
for every index $i$. Above, $ \bar\sigma_{i}$ denotes a positive constant and $\xi$ some deterministic function such that $\lim_{t\rightarrow T}
\int_0^t \xi_{u}^2 \, du = +\infty$.
\item[(ii)] there exist some vectors $\rho_i^S$ in $\RR^p$ such that
\begin{equation}
 \rho_{ij} = E_P[ dW_{it}.dW_{jt}]/dt  = \left[\frac{\bar\sigma_{i}^2 + \bar\sigma_{j}^2}{2\bar\sigma_{i}\bar\sigma_{j}}\right](\rho_i^S)'\rho_j^S, \;\; \forall i\neq j.
\label{GenCondExistenceBEC}
\end{equation}
\item[(iii)] for every index $i$, the correlations that are
used for the pricing of the structured product, in the $p$-factor Gaussian copula model, are given by the relations $ \rho_{i} = \rho_i^{S}$, $i=1,\ldots,n$.
\end{enumerate}
\end{theorem}

\begin{corollary}
Under the assumptions of theorem~\ref{GenUnicity}, assume that the matrix of spread correlations $\Sigma^S$ is $p$-factor, i.e.
there exist $p$-dimensional vectors $\rho_i^S$ such that $\rho_{ij}= (\rho_i^S)'\rho_j^S$ for all $i\neq j$.
Then, if the coefficients $\bar\sigma_{i}$ are the same for every index $i$, the pricing by replication under (pCop) is possible.
In this case, $\rho_i = \rho_i^S$.
\end{corollary}

The result above can be rewritten in a more striking way. If the
matrix $\tilde\Sigma^S=[\{\frac{2\bar\sigma_{i}\bar\sigma_{j}}{\bar\sigma_{i}^2 +
\bar\sigma_{j}^2}\}\rho_{ij}]$ is
given by a $p$-factor correlation structure, then the Gaussian copula
pricing model should be $p$-factor. And the corresponding pricing
correlations are given by the decomposition of $\tilde\Sigma^S$, that is a correlation matrix (see section~\ref{StructuralModels}).
This is clearly in contrast with the current practice, where
practitioners use only one-factor Gaussian copula model by far.

\section{Analysis of a First $p$-th to default}
\label{FPTD}
The simplest basket credit derivative is a First-To-Default, where the payoff is non zero, except when all the names have
survived until the maturity date $T$. By assuming that all the names share the same recovery rate $R$ and with our previous notations, it means
$$ \psi_{FtD} (\bdelta ) = (1-R).\left(1- \prod_{j=1}^n (1-\delta_j)\right).$$
Invoking theorem~\ref{theocopG}, between $t$ and $t+dt$ and under
the one factor copula model, the infinitesimal drift variation of a
First-To-Default value is
\begin{equation}
d\mu_{t,FtD} := \frac{(1-R)dt}{2(1-\rho^2)}
\sum_{i,j,i< j}^n [ (\beta_{it}^2+\beta_{jt}^2)\rho_i \rho_j - 2\beta_{it} \beta_{jt} \rho_{ij} ]
\int \Prod_{k\neq i,j} q_{k|x}. \phi(d_i)\phi(d_j)\phi(x) \, dx,
\label{FTDpi}
\end{equation}
if no default occurs in this time interval. Above, we recover the
main formula of~\cite{StratReport}. Assume that all the correlation
levels are the same, in the real world and for pricing purpose: all
the spread correlations $\rho_{ij}$ are equal and are denoted by
$\rho^2_S$, and $\rho_i:=\rho$ for all $i$. It is important to note
that, whatever the values of correlations and spreads:
\begin{itemize}
\item when $\rho=0$, $d\mu_{t,FtD} < 0$.
\item when $\rho=\rho_{S}$, $d\mu_{t,FtD} > 0$.
\end{itemize}
Thus, the break-even correlation of a First-To-Default lies somewhere between zero and the corresponding (uniform) spread correlation.

\medskip

Actually, it is possible to find a general formula to deal with First $p$-th to default, $p=1,\ldots,n$.
The default leg of such securities is the amount of loss in the portfolio, but capped at the $p$-th default.
In other words, the payoff of a First $p$-th to default is
$$ \psi_p = \min(\sum_{i=1}^n \delta_i,p).(1-R).$$
Such securities are important. Indeed, they may be seen as basic
building blocks for most standard basket credit derivatives,
including CDO tranches (see Brasch 2006). Now, we calculate the
break-even correlations of First $p$-th to default explicitly.

\begin{theorem}
\label{FirstpTD}
Under the assumptions (Cop) (I) (Q) (F1) (A), the value of a First $p$-th to default
 will change between $t$ and $t+dt$ by the amount
\begin{align*}
dV^{(p)}_t &=(\cdots)\cdot d\vec{W}_t+
\frac{(1-R)dt}{2} \sum_{i<j}
\left[ \rho_i \rho_j \beta_{it}^2  + \rho_i \rho_j \beta_{jt}^2-2 \beta_{it} \beta_{jt} \rho_{ij}  \right]
\cdot A_{ij}^\ast,
\end{align*}
where
$$ \beta_{it} := \frac{\sigma_{it}}{\phi(\Phi^{-1}(Q_{it})) },\;\;\;A_{ij}^\ast:=\sum_{\substack{\delta_k; k\leq n, k \neq i,j \\ \sum_k \delta_k = p-1}} A_{ij}(\bdelta)$$
and
$$ A_{ij}(\bdelta):= \int \Prod_{k\neq i,j} p_{k|x}^{\delta_k}q_{k|x}^{1-\delta_k}
 \frac{\phi(d_i)\phi(d_j) \phi(x)}{\sqrt{1-\rho_i^2}\sqrt{1-\rho_j^2}}\, dx.$$
\end{theorem}
See the proof in the appendix~\ref{DemFPTD}. Check that we recover
equation~(\ref{FTDpi}) when $p=1$. The previous reasoning we have
led on the First-To-Default is available for a general First $p$-th
to default: in the case of identical spread correlations $\rho^2_S$
and identical pricing correlations, the break-even correlation level
lies between zero and $\rho_S^2$. This result is true for arbitrary
values of spread volatilities and every initial credit spread
curves.

\medskip

At first glance, the theorem~\ref{FirstpTD} allows to write the Break-even correlation as a linear combination of spread correlations,
 with positive coefficients. Indeed, since all the quantities $A_{ij}^\ast$ above are positive, we can rewrite
$$ \rho_{BE}^2 = \frac{2\sum_{i< j}A_{ij}^\ast\beta_{it}\beta_{jt} \rho_{ij}}{\sum_{i < j} A_{ij}^\ast(\beta_{it}^2 + \beta_{jt}^2)}
:= \sum_{i\neq j}w_{ijt} \rho_{ij},$$
where all the $w_{ijt}$ belong to $[0,1]$. Nonetheless, the previous coefficients $w_{ijt}$ depend on $\rho_{BE}$ itself, through
the so-called $d_i$ terms. Moreover, the $w_{ijt}$ depend on the current values of the survival probabilities $Q_{it}$ and $Q_{jt}$, i.e. they are time-dependent.
Thus, the calculation of the instantaneous break-even correlation of a First $p$-th to default implies solving
an implicit equation.

\medskip

If all the $Q_{it}$ are equal, $i=1,\ldots,n$, then the previous terms $A_{ij}^\ast$ are equal. Thus, in this case, the
(instantaneous) break-even correlation is given by
\begin{equation}
\rho_{BE}^2 = \frac{2}{\sum_{i < j} (\beta_{it}^2 + \beta_{jt}^2)}\cdot\sum_{i< j}\beta_{it}\beta_{jt} \rho_{ij}.
\label{BECorrWeighted}
\end{equation}
Note that this equation is true formally when there are two names
only in the basket. In the latter equation, the break-even
correlation depends explicitly on empirical correlations $\rho_{ij}$
and some ``volatility-type'' coefficients $\beta_i$, that are equal
to the previous quantities $\bar\sigma_{i}\xi_t$ (see
section~\ref{GCMrevisited}). For instance, if all the spread
volatilities and survival probabilities are the same, then
$$\rho_{BE}^2 = \frac{2}{n(n-1)}\cdot\sum_{i< j}\rho_{ij}.$$

\section{A bridge towards structural models}
\label{StructuralModels}
All the previous analysis has been led
inside the family of the copula based models. Traditionally,
these models are defined through a direct specification of the joint
laws of default times, but here, we have just needed the joint law
of default events at the time horizon $T$. Now, it is fruitful to
link this framework to structural models, following the seminal
paper of Merton (1974). Classically, they are defined in terms of
asset value processes $(A_{it})$ that follow some particular
dynamics in continuous time. The default time of a name $i$ is
defined as a functional of $i$'s asset value trajectory.

\medskip

The simplest case is given by the usual one-period Merton model, where
\begin{equation}
 (\tau_i > T ) := (A_{iT} \leq b_{iT}),
\label{DefDef}
\end{equation}
for deterministic boundaries $b_{iT}$ that are calibrated to be
consistent with current observed default probabilities. Since the
knowledge of asset values is equivalent of the knowledge of default
likelihoods (or equivalently credit spreads), our asset values
should depend on the same drivers as the survival probabilities
$Q_{it}$, i.e. on the previous Brownian motions $(W_{it})$,
$i=1,\ldots,n$. Typically, it is assumed the (risk-neutral) asset
values follow some Ito processes
$$ dA_{it} = \tau_i(t,A_{it}) dW_{it}  + \nu_{i}(t,A_{it}) dt,$$
where $(W_{it})$ is a brownian motions under $Q$. For some classical
families of diffusion processes~\footnote{geometric Brownian motion
or Ornstein-Uhlenbeck process, e.g.}, the survival events can be
rewritten
\begin{equation}
(A_{iT} \leq  b_{iT})=\left(\int_0^T \psi_i(u) dW_{iu} \leq  \bar b_{iT}\right),
\label{DefDef2}
\end{equation}
for some deterministic function $\psi_i$ and another default
boundary $\bar b_{iT}$. Note that the thresholds $b_{iT}$ and
$\bar{b}_{iT}$, $i=1,\ldots,n$, do not depend on $t$. In this
section, we assume we can write~(\ref{DefDef2}). We deduce the
survival probabilities that are implied by this structural approach:
\begin{equation}
Q_{it} (T)= Q(A_{iT} \leq  b_{iT}|\F_t) = \Phi \left( \frac{ \bar b_{iT} - \int_0^t \psi_i(u) dW_{iu}}{ \sqrt{ \int_t^T \psi_i^2} } \right) .
\label{QMerton}
\end{equation}

\medskip

Now, in the previous Merton framework, assume that the correlation
structure among the underlying names is one-factor:
$$ W_{it} = \theta_i W_t + \sqrt{1-\theta_i^2} W_{it}^*,\;\;\; i=1,\ldots,n,$$
with independent Brownian motions $(W_{it}^*)$ under $Q$, and a common $Q$-Brownian motion $(W_t)$ that generates some
dependency between the underlying default times. Thus, the default events $(\tau_i\leq T)$ are
independent given the r.v. $Z_i:=\int_0^T \psi_i(u) \, dW_u/ (\int_0^T \psi_i)^{1/2} $, $i=1,\ldots,n$.
If the r.v. $Z_i$ are not the same for different names $i$, the pricing procedure
involves multiple integrals, and can become tedious in practice. Hopefully, the
usual Merton model is defined by $\psi_i=1$, and in this case, we
recover the usual one-factor Gaussian copula specification.
Otherwise, a Merton model with general functions $\psi_i$ will
induce $m$-factor models, $m\geq 1$ and $m\leq n$.

\medskip

Therefore, by defining default events through the relations
\begin{equation}
(\tau_i > T) = \left(\int_0^T \psi_i(u) dW_{iu} \leq  \bar b_{iT}\right),
\label{DefDefaultMerton}
\end{equation}
we get back the Gaussian copula model, possibly in a multi-factor version.

\medskip

Obviously, the available information at time $t\leq T$ is the
current survival probabilities $(Q_{1t}(T),\ldots,Q_{nt}(T))$, or
equivalently the current asset values, or even $(\int_0^t
\psi_1(u)\, dW_{1u},\ldots, \int_0^t \psi_n(u)\, dW_{nu})$. Thus, we
can write $\F_t = \sigma\left(\int_0^v \psi_i(u)\, dW_{iu},
i=1,\ldots,n;v\leq t \right)$. Moreover, the common factor at time
$t$ (the ``famous'' $X$) is given by the trajectory of the common
brownian motion $(W_t)$ between $t$ and $T$. To be specific, $X$ can
be reduced to the knowledge of the random variables $\int_t^T
\psi_i(u)\, dW_{u}$, $i=1,\ldots,n$.

\medskip

Thus, let us define the filtration $\Gc=(\Gc_t)_{t\in [0,T]}$,
$\Gc_t:=\sigma(\int_v^T \psi_i(u)\, dW_{u}, i=1,\ldots,n; v\leq t)$,
and the extended filtration $\Hc=(\F_t \vee \Gc_t)_{t\in [0,T]}$.

\medskip

For instance, in the standard gaussian copula model, the common
factor $X$ at time $t$ is univariate and is simply $W_T - W_t$, the
increment of the common brownian motion between $t$ and $T$. In this
case, $\psi_i=1$ for every $i$. We prove in the
appendix~\ref{ProofMarting}:

\begin{theorem}
In the 1FGCM, the conditional probabilities $(p_{jt|x=W_T-W_t})_{t\in [0,T]}$, as given in equation~(\ref{GCMSpecif}), are $\Hc$-martingale under $Q$.
\label{MartingaleCondProb}
\end{theorem}

Now, we are interested in the $Q_{it}$-dynamics that are implied by
the previous Merton model. We would like to compare them with the
dynamics we have been obtained in theorems~\ref{Unicity}
and~\ref{GenUnicity}, through fully different arguments.
From~(\ref{QMerton}), we deduce the dynamics of these survival
probabilities under $Q$:
\begin{equation}
dQ_{it} =  -\beta_{it} \phi(\Phi^{-1}(Q_{it})) dW_{it},\;\; \beta_{it} :=\frac{\psi_i(t)}{\sqrt{ \int_t^T \psi_i^2}} \cdot
\label{DynSpreadMerton}
\end{equation}
Note that $\beta_{it}$ is not random. Therefore, when $t_0<t$, we get easily
$$\int_{t_0}^t \beta^2_{iu} \,du = \ln(\int_{t_0}^T \psi_i^2 ) - \ln(\int_{t}^T \psi_i^2 ) , $$
that tends to $+\infty$ when $t$ tends to $T$.
Actually, there is a one-to-one mapping between the deterministic functions
$\beta_{i}$ and $\psi_i$ because
\begin{equation}
 \psi_i(t) :=C \beta_{it} \exp\left( -\int_{0}^t \frac{\beta_{iu}^2}{2} \, du \right),
\label{MertonOpposite}
\end{equation}
for some arbitrary positive constant $C$.

\medskip

Therefore, if the latter functions are of the
type $\beta_{it}=\bar\sigma_i\zeta_t$, or equivalently
\begin{equation}
 \psi_i(t) :=C_i \zeta_t \exp\left( -\bar\sigma_i^2\int_{0}^t \frac{\zeta_{u}^2}{2} \, du \right),
\label{MertonOpposite2}
\end{equation}
then there is identity between the spread
dynamics induced by the Merton approach (see~(\ref{DefDefaultMerton}) and~(\ref{DynSpreadMerton})) and those obtained in theorem~\ref{GenUnicity} by reverse engineering the GCM.

\medskip

For example, if the underlying asset values follow Brownian motions,
then $A_{it} :=  W_{it}$ and, with the previous notations,
$\psi_i(u)=1$ and $ b_{iT} =\bar b_{iT}= \sqrt{T}
\Phi^{-1}(Q_{i0}(T))$. This is the most usual specification of a
one-period Merton model. Then, in this case, $\beta_{it} = \zeta_t=
1/\sqrt{T-t}$ for all the names. With this very simple
specification, the spread dynamics do not involve any firm-specific
``volatility-type'' coefficient $\bar\sigma_i$: $\bar\sigma_i=1$ for
all $i$.

\medskip

Actually, the latter couple of functions $(\beta_{it}, \psi_{it})$ is not the single one.
For instance, we could propose
\begin{equation}
(\beta_{it}, \psi_{it}) :=
\left(\frac{\bar\sigma_i}{\sqrt{1-t/T}},\bar\psi_i\bar\sigma_i
(1-t/T)^{\bar\sigma_i^2 T/2-1/2}\right). \label{MertonExt}
\end{equation}
Note that this choice induces almost the same dynamics as in the
previous case. But now, there is more heterogeneity in terms of
spread dynamics, through different coefficients $\bar\sigma_i$.
Therefore, to get heterogeneity (or ``volatility'') effects, we can
define the underlying survival events by
$$ (\tau_i>T) := \left( \int_0^T (T-u)^{[\bar\sigma_i^2 T-1]/2}\, dW_{iu} \leq \bar{b}_{iT}   \right).$$
Nonetheless, to price a basket in this model, it is necessary to integrate the conditional default probabilities with respect to several factors.

\medskip

In the specification~(\ref{MertonExt}), it is nice to recover the
results of section~\ref{GCMrevisitedExtended}. With the notations of
the latter section, the correlation between survival probabilities
(or ``spread'' moves) $\rho_{ij}$ is the correlation between the
previous Brownian motions $(W_{it})$~\footnote{see
equation~(\ref{DynSpreadMerton})}. Moreover, the correlations
between the pricing factors were denoted by $\rho'_i\rho_j$ when
$i\neq j$~\footnote{recall equation~(\ref{GenGCMSpecif}). The asset
value processes are no more one-factor, but now $p$ factors, $p\geq
1$, due to different functions $\psi_i$.}. Since the asset values at
maturity are given by $A_{iT}= \int_0^T \psi_{it}\, dW_{it}$,
$i=1,\ldots,n$, by definition of the vectors $\rho_i$, we should
satisfy $\rho'_i\rho_j = Corr(A_{iT},A_{jT})$. But simple
calculations provide
\begin{equation}
Corr(A_{iT},A_{jT}) = \frac{\rho_{ij} \int_0^T \psi_i \psi_j}{\sqrt{\int_0^T \psi_i^2}\sqrt{\int_0^T \psi_j^2} }
= \rho_{ij} \frac{2\bar\sigma_i \bar\sigma_j }{\bar\sigma_i^2+ \bar\sigma_j^2} \cdot
\label{CorrelAssetValuesMerton}
\end{equation}
Thus, we recover the key relation of theorem~\ref{GenUnicity}:
\begin{equation}
\rho'_i\rho_j = \rho_{ij} \frac{2\bar\sigma_i \bar\sigma_j }{\bar\sigma_i^2+ \bar\sigma_j^2} \cdot
\label{CorrelAssetValuesMerton2}
\end{equation}
Note that, through the equation~(\ref{CorrelAssetValuesMerton}), we have proved that, for an arbitrary
correlation matrix $[\rho_{ij}]$, every symmetrical matrix of the type
$$ \left[ \rho_{ij} \frac{2\bar\sigma_i \bar\sigma_j }{\bar\sigma_i^2+ \bar\sigma_j^2}\right]$$
is a correlation matrix. This was far from obvious.

\medskip

Another model would be to set
$$(\beta_{it}^{(\alpha)}, \psi_{it}^{(\alpha)}) := \left(\frac{\bar\sigma_i}{(1-t/T)^{\alpha}},\frac{\bar\psi_i\bar\sigma_i }{(1-t/T)^{\alpha}}
\exp\left(- \frac{\bar\sigma_i^2 T}{2(2\alpha -
1)}[(1-t/T)^{1-2\alpha}-1]\right) \right),$$ for some constants
$\alpha < 0.5$, $\bar\sigma_i>0$ and $\bar\psi_i>0$. Under the
latter Merton-type specifications, the asset value processes are
different from usual Brownian motions, but follow (no drift, for
instance) $ dA_{it} = \psi_{it}^{(\alpha)} dW_{it}.$
Even in this
case, we are still able to exhibit spread dynamics that allow
perfect replication of CDO tranches. Through
theorem~\ref{GenUnicity}, these are
\begin{equation}
dQ_{it} =  \beta_{it}^{(\alpha)} \phi(\Phi^{-1}(Q_{it})) dW_{it} +\mu_{it} dt.
\end{equation}
Simple calculations show that, again with the latter specification,
we recover the usual relation~(\ref{CorrelAssetValuesMerton2})
between spread correlations, pricing correlations and the
``volatilities'' $\bar\sigma_i$.

\medskip

We have restricted our analysis to default events that are induced
by structural models but through the relation~(\ref{DefDef}), or,
equivalently, through the relation~(\ref{DefDef2}). Other choices
would be possible, for instance by defining a default event as a
first hitting time of a boundary, i.e. $(\tau_i >T) =( A_{it}<
b_{iT} , \;\; \forall t \in [0,T] )$. An open question would be to
study such alternative structural models, and to exhibit the spread
dynamics that would be consistent with a ``perfect'' replication.
Since multivariate extensions of First-hitting time models are very
complicated in analytic terms (Zhou 2001, Hull et al. 2006), this
avenue is left for further research.

\medskip

It is interesting to note that the previous analysis has some
striking applications in equity derivatives. A European digital put
on $i$'s stock has a payoff $(S_{iT} \leq K)$, for a given strike
$K$ and a maturity $T$. If we consider this stock follows a
lognormal diffusion with constant volatilities $\sigma_i$ and a
constant short rate $r$, then the $t$-price of this security is then
$$P_{it} := Q(S_{iT} \leq K|\Fc_t)= \Phi\left(\frac{1}{\sigma_i\sqrt{T-t}} \left[ \ln(\frac{K}{S_{it}}) -
(r-\frac{\sigma_i^2}{2}).(T-t)\right]    \right) .    $$ The
dynamics of this price is the same as the classical one-period
Merton model (when the asset values follow Brownian motions):
$$ dP_{it} = \frac{1}{\sqrt{T-t}}\phi\left(\Phi^{-1}(P_{it})
\right)\, dW_{it}.$$ Moreover, the payoff of a digital worst-of
European equity options is
$$V_{T,dwo} = {\bf 1}(S_{iT} \leq K_i, \mbox{ for
some } i \in \{1,\ldots,n\} ).$$ It is similar to the payoff of a
First-to-default, after having defined the default events by
$(A_{iT} \leq b_{iT})$, $i=1,\ldots,n$. Thus, the pricing formula of
a First-to-default and the replication results we got in
section~\ref{FPTD} can be applied to the case of worst-of put
options. For instance,  at time $t$, the price of the worst-of put
option on $n$ names is \begin{equation} V_{t,dwo} = 1-\int_{\RR^p}
\prod_{i=1}^n (1-\Phi)\left(\frac{\Phi^{-1}(P_{it}) - \rho_i'
\xx}{\sqrt{1-|\rho_i|^2}} \right)\, \phi(\xx)\, d\xx,
\label{pricingWO}
\end{equation}
by assuming the correlations between stock processes are given by a
$p$-factor model. Table~\ref{MyMap} summarizes the correspondence
between the credit derivative and the equity derivative frameworks.
If the true (historical) processes of stocks are lognormal
diffusions, then historical correlations between these stock
returns~\footnote{or equivalently between the put prices $P_{it}$}
$\rho_{ij}$ should be invoked in the pricing
formula~(\ref{pricingWO}). Nonetheless, it is not the case and the
pricing formula~(\ref{pricingWO}) is still used, then these
correlations should be different. For instance, if the stock returns
follow diffusions
\begin{equation}
d(\ln S_{it})=\int_0^t \psi_i(u) \, dW_{iu} + \mu_{it} \,dt,
\label{dynStockPsi}
\end{equation}
for some deterministic functions $\psi_i(t)=\bar\sigma_i T
(1-t/T)^{[\bar\sigma^2 -1]/2}$ as above, then the pricing factors
$\rho_i$ that will be put in formula~(\ref{pricingWO}) are now given
by the relations
$$ \rho_i'\rho_j = \frac{2\bar\sigma_i\bar\sigma_j}{\bar\sigma_i^2
+\bar\sigma_j^2} \rho_{ij}.$$ Nonetheless, a stock dynamics as
in~(\ref{dynStockPsi}), even if not impossible, is unlikely because
it depends on the option maturity $T$. Thus, a bootstrap-type
procedure would be necessary to build the whole dynamics of stocks.

\begin{table}
\begin{center}
\begin{tabular}{||l|c|c||}
\hline\hline  & Credit world & Equity world  \\ \hline
Underlying & asset value & stock  \\
Vanilla derivative & default probability & digital put  \\
Structured derivative & First-to-default & worst-of put  \\
\hline\hline
\end{tabular}
\caption{One-to-one mapping between the European credit derivatives
and the European equity derivatives. \label{MyMap}}
\end{center}
\end{table}

\section{Alternative models}
\label{AlternativeModels}
In section~\ref{GCMrevisited}, we have proved that the Gaussian copula model can be considered (under some conditions) as a replication model if the individual survival
probabilities follow some diffusion-type processes (theorem~\ref{theocopG}). A natural question would be to find
a similar result for other credit risk models, i.e. for alternative specifications of the quantities $p_{i|x}$.
Surprisingly, it is absolutely not easy. Particularly, no Archimedean copula model can be seen as
a replication model, in our framework, as it is specified in the next theorem. Recall that a usual Archimedean (static) model is defined
by a one-dimensional non negative random variable $X$, whose Laplace transform is denoted by $\xi$, and by the
conditional default probabilities

\begin{equation}
 p_{i|x} := \exp\left(-x \bar \xi_i \right), \;\;\; \bar \xi_i := \xi^{-1}(1-Q_{it}(T)),
\label{ModelArch}
\end{equation}
for all $x\geq 0$.

\begin{theorem}
\label{Archim}
For any Archimedean copula model~(\ref{ModelArch}), assume that
\begin{itemize}
\item $E[X^2]<+\infty$
\item the volatility functions $\sigma_{it}$ do not depend on $i$ except through the trajectories
$Q_{iu}$, $u\leq t$.
\item The ``real world'' correlation levels $\rho_{ij}$ at time $t$ do not depend
on the survival probabilities $Q_{ku}$, $u\leq t$, $k\in \{1,\ldots,n\}$.
\end{itemize}
Then there are no $Q_{it}$-dynamics of type (Q) such that the price process $(V_t)$ of any payoff $\psi$ is a $\F$-martingale under $Q$.
\end{theorem}

Actually, if we allow the spread correlations $\rho_{ij}$ to be time-dependent, the picture is different.
Indeed, some of the previous results can be extended to random correlation levels in a straightforward way. To be specific, we
assume now that the $t$-instantaneous correlation level $\rho_{ij}$ between $dW_{it}$ and $dW_{jt}$ can depend on the past realized values
of the survival probabilities $Q_{ku}$, $k\in \{1,\ldots,n\}$, $u\leq t$. The main theorem~\ref{thdpit2} is still true because it is related to hedging strategies between two immediate successive times $t$ and $t+dt$ and conditional on past information $\Fc_t$.

\medskip

Under this extended framework, we are able to exhibit an alternative
to the Gaussian copula model, i.e. a ``static'' model that has an
interpretation in terms of replication and the underlying
single-name dynamics. The Clayton copula model provides such an
alternative, even if it is an Archimedean model.

\medskip

We recall that, in the Clayton copula model, there exists a single common factor $X$ that follows a Gamma law
with the density $f_X(x)=x^{1/\theta -1} \exp(-x) /\Gamma(1/\theta)$ on $\RR^+$, for some nonnegative parameter $\theta$.
At time $t$, the default probability of $i$ before $T$ is given by
\begin{equation}
p_{i|x}(T)=\exp\left(-x[(1-Q_{it}(T))^{-\theta} -1]\right).
\label{ClaytonModel}
\end{equation}
In this model, the key dependence parameter is now $\theta$, that is independent of $i$. Thus, the
heterogeneity between names is driven here by the heterogeneity in terms of survival probabilities
only. Now, we change slightly the dynamics of the $Q_{it}$:

\medskip

Assumption (Q'): Under the risk neutral measure $Q$, the survival probability process of name $i$ is given by
\begin{equation}
dQ_{it}(T) = \sigma_{it} dV_{it},\;\;t\leq T,
\label{processQ2}
\end{equation}
when $i$ is not defaulted. Here $(\sigma_{it})$ are general ${\mathcal F}$-adapted process, and
$$ dV_{it}:=\beta_i(Q_{it}) \, dZ_t  + \sqrt{ 1- \beta^2_i(Q_{it})} \, dZ_{it}^\ast,\;\; \beta_i(Q) := \left(1-(1-Q)^{\theta}\right)^{1/2},$$
where $Z$ and $Z_i^\ast$, $i=1,\ldots,n$ denote $n+1$ independent $\F$-Brownian motions under $Q$.

\medskip

The processes $(V_{it})$ are no more Brownian motions, in general.
The following result is proved in the appendix.
\begin{theorem}
\label{Clayton}
Under the Clayton copula model, assume (I) (Q') (F1) (F2) (A). If
$$\sigma_{it}=\sigma_0 (1-Q_{it}).\left(1-(1-Q_{it})^{\theta} \right)^{1/2},\;\; i=1,\ldots,n,$$
then the pricing formulas given by the Clayton model are
$\F$-martingales under $Q$.
\end{theorem}

Thus, we have exhibited new spread dynamics that are consistent with
the replication of CDO tranches, when they are priced with the
Clayton copula. Moreover, we have linked the (unique) model
parameter $\theta$ with the true spread process, through the
instantaneous correlations between future spread moves. Note that,
in theorem~\ref{Clayton}, when the default probabilities of name
$i$ goes to one, its spread correlations tend to zero. This is
in line with intuition: all other things being equal, this name becomes more
idiosyncratic.

\medskip

Moreover, under the pricing point of view, the previous Clayton
model can be interpreted as a structural model one, like the
Gaussian copula can be linked to the one-period Merton model. In the
latter 1FGCM case, the underlying asset values are driven by
Brownian motions in the 1FGCM. When dealing with the Clayton copula,
we can consider asset values that are driven by a common Gamma
process $Y$. Recall that, by definition,
\begin{itemize}
\item $Y(0)=0$,
\item $Y$ has independent increments,
\item $Y(t)-Y(s)$ follows a Gamma distribution $\Gamma(\gamma.(t-s),\lambda)$, $t>s$.
\end{itemize}
Recall that a random variable that follows a law
$\Gamma(\gamma,\lambda)$ has got the density $x\mapsto
\lambda^\gamma\exp(-\lambda x) x^{\gamma -1}/\Gamma(\gamma) $ on
$\RR^+$. Its Laplace transform is
$$ E[\exp(-s X)]= \left( 1+ \frac{s}{\lambda} \right)^{-\gamma}.$$
In~\cite{Singpur} (rediscovered recently in~\cite{Joshi}), a
multivariate credit model can be defined by assuming the integrated
hazard function of every name (say $i$) is given by
$$ \Lambda_i(t) = \int_0^t c_i(u) \, Y(du),$$
for a given deterministic positive function $c_i$. In other words,
at time $0$, the default times are defined as
$$ \tau_i = \inf\{t \geq 0 | \int_0^t c_i(u) \, Y(du) > E_i \},$$
where $E_1,\ldots,E_n$ denote independent exponentially distributed
random variables. Equivalently, we can think in terms of asset
values $X_i$, that would be defined by
$$ X_i(t)= \phi_t\left( -\frac{\ln(U_i)}{Y(t)}  \right), \;\; \phi_t(s) := E[\exp(-s Y_t)].$$
Here, $U_1,\ldots,U_n$ denote independent random variables that are
uniform on $[0,1]$. In a one-period version of the associated
structural model, the default event at time horizon $T$ is defined
by
$$ (\tau_i \leq T ) =( X_i(T) \leq b_i(T)),$$
for some boundary $b_i(T)$. With this specification, the common
factor is the random variable $Y(T)$, and we can check that
$b_i(T)=1-Q_{i0}(T)$. Thus, we deduce
$$ P(\tau_i \leq T | Y(T)=x,\Fc_0)=\exp\left( -x\lambda [(1-Q_{i0}(T))^{-1/\gamma T} -1 ]\right).$$
And we retrieve the equation~(\ref{ClaytonModel}) when $\lambda=1$
and $ \theta\gamma T=1$.

\medskip

In another perspective, we can try to find alternative replication
models where every individual conditional probability is of the type
\begin{equation}
p_{i|x} := \eta\left(\alpha_i x - h(Q_{it}) \right),
\label{piGCMExtended}
\end{equation}
for some functions $\eta$, $h$ and some constants $\alpha_i$. The
density of the common factor $X$ is unknown, is denoted by $f_X$ as
usual and its support is $\RR$. Is is not a lack of generality: we
could replace $x$ by a function of $x$ in~(\ref{piGCMExtended}), and
changing $X$'s density conveniently.

\medskip

Note that the Gaussian copula model belongs to the latter family of
specifications. It is the case of the Gamma-pool model too
(\cite{Garcia},~\cite{Jaeckel}), but not of Archimedean models.
Thus, it is tempting to search some models of the
type~(\ref{piGCMExtended}) that can be seen as replication models,
i.e. that satisfy equation~(\ref{GeneralCondition}). Actually, it
can be proved that the Gaussian copula only satisfies this
requirement.

\begin{theorem}
\label{GCMTheTrue} Consider a model of the
type~(\ref{piGCMExtended}), with unknown functions $\eta$, $h$ and
$f_X$. Assume that the real world correlations $\rho_{ij}$ are
constant and that the volatilities $\sigma_{it}$ depend on $i$ only
through the survival probabilities $Q_{i,.}$. This model satisfies
equation~(\ref{GeneralCondition}) if and only if it is the Gaussian
copula model, i.e. iff it is defined through
equation~(\ref{GCMSpecif}).
\end{theorem}
See the proof in the appendix.

\medskip

Thus, the difficulty in finding alternative models can be seen as an
argument in favor of the Gaussian copula model. In other words, we
can say that this specification has some nice analytical properties,
that cannot be found elsewhere easily. The latter point should be
stressed, when a lot of people criticize the Gaussian copula model
(see Lipton and Rennie 2008, for instance). To summarize, the
one-factor Gaussian copula model seems to be the most achieved one
at least amongst of static models. Indeed, for this model only, we
are able to
\begin{itemize}
\item exhibit spread dynamics that are consistent with the model specification,
\item make an explicit link between the pricing parameters and the parameters in the real-world dynamics,
\item price by replication when no jump-to-default occurs.
\end{itemize}

\medskip

To understand why the Gaussian copula model is so natural can be
understood by going back to structural models. If the underlying
asset values $A_{it}$ are driven by some Brownian motions, it is
tempting to define the survival of $i$ at time $T$ as the event
"$A_{iT}$ is larger than some fixed constant". For most "natural"
specifications, this is equivalent to "$W_{iT}$ is larger than some
fixed constant". In the latter case, we retrieve the Gaussian copula
framework. To get alternatives, one way would be to leave the usual
Brownian world, especially concerning the process that is followed
by the survival probabilities $Q_{it}$. For instance, by assuming
that the asset values (and thus the survival probabilities) are
Levy-driven processes, but we have checked in
theorem~\ref{GCMTheTrue} that the Gamma-pool model does not satisfy
equation~(\ref{GCMSpecif}), unfortunately. Since these Levy-driven
models allow spread jumps, the market efficiency can no more be
obtained without considering alternative hedging strategies and
hedging instruments. Another less natural way would be to define
asset values still through usual diffusion processes, but by
defining the default events as $(A_{iT} \in \Ic_i)$ for some unusual
real subset $\Ic_i$, or even as functions of the whole random
trajectories $(A_{it})$, $t\in [0,T]$.

\section{Empirical analysis of break-even correlations}
\label{Empiric}

To illustrate these results, consider first a simple example: a
First-to-default, second-to-default and third-to-default securities
that are written on four names. We assume that the survival
probabilities in the real world follow the "correct" theoretical
dynamics~(\ref{BEDynamics}) without any drift. To simplify, we
assume all the volatilities $\bar\sigma$ are the same constant
functions. Moreover, we assume that the dynamics associated to the
first two names (say $1$ and $2$) are independent from the ones of
the two other names ($3$ and $4$). The "spread" correlation between
the first two names is $\rho_{12}^S=30\%$, when it is
$\rho_{34}^S=70\%$ between the two other names.

\medskip

It is relatively easy to check that the theoretical instantaneous
break-even correlations for all the First $p$ to default are given
by
$$ \rho_{BE}^2 = \frac{A_{12}^\ast \rho_{12}^S +  A_{34}^\ast\rho_{34}^S}{A_{12}^\ast+A_{13}^\ast+A_{14}^\ast+A_{23}^\ast+A_{24}^\ast+A_{34}^\ast}, $$
with the notations of theorem~\ref{FirstpTD}. Obviously, the
coefficients $A_{jk}^\ast$ depend on the order of the security, i.e.
on $p$ and on the current survival probabilities (or intensity
levels) on every particular trajectory that is simulated. We draw
some $Q_{it}$, $i=1,\ldots,4$, trajectories on $180$ daily time
intervals. Spread volatility coefficients $\bar\sigma_t$ are assumed
to be constant in time and the same per names. They take the value
$50\%$. The first two names (resp. the last two names) share the
same spot intensity level. Therefore, by averaging overs $100$
random trajectories, we are able to calculate mean break-even
correlations on the whole time interval.

\medskip

If the spread trajectories are similar between all the names, then
all the coefficients $A_{jk}^\ast$ are similar and the instantaneous
break-even correlation is around $(\rho_{12}^S + \rho_{34}^S)/6
\simeq 16.7 \%$. Note that this level is independent from the strike
we consider. Moreover, when the spreads of names $1$ and $2$ are a
lot smaller than those of names $3$ and $4$~\footnote{on the whole
trajectory we consider}, then we can check easily on the theoretical
expression~(\ref{BEDynamics}) that
\begin{itemize}
\item   for the First-To-Default, $A_{12}^\ast << A_{34}^\ast$ and $\rho_{BE}^2 \simeq \rho_{34}^S = 70\%$, and
\item for the Third-To-Default, $A_{34}^\ast << A_{12}^\ast$ and $\rho_{BE}^2 \simeq \rho_{12}^S=30\%$.
\end{itemize}
Indeed, these theoretical relations are observed in our simulation
study: see tables~\ref{FTDTable},~\ref{F2TDTable}
and~\ref{F3TDTable}.

\begin{table}
\begin{center}
\begin{tabular}{||lll||c|c|c|c||}
\hline\hline \multicolumn{7}{||r||}{Spot Default intensities of
$1$ and $2$} \\ \hline & & & $0.1\%$ & $1\%$ & $5\%$ & $30\%$
\\ \hline
\multirow{5}{*}{\begin{sideways} Intensities \end{sideways}} &
\multirow{5}{*}{\begin{sideways}  $3$
and $4$\end{sideways}} & $0.1\%$ & {\it 18}  & 20     & 26 & {\bf 30}   \\
& & $1\%$ & 45 & {\it 17}  & 17 & 27  \\
& & $5\%$ & 68 & 39  & {\it 16} & 18  \\
& & $30\%$ & {\bf 69}  & 69  & 53 & {\it 16}  \\
\hline\hline
\end{tabular}
\caption{Break-even correlations of the First-To-Default in the
four-name basket ($\bar\sigma=50\%$). \label{FTDTable}}
\end{center}
\end{table}

\begin{table}
\begin{center}
\begin{tabular}{||lll||c|c|c|c||}
\hline\hline \multicolumn{7}{||r||}{Spot Default intensities of
$1$ and $2$} \\ \hline & & & $0.1\%$ & $1\%$ & $5\%$ & $30\%$
\\ \hline
\multirow{5}{*}{\begin{sideways} Intensities \end{sideways}} &
\multirow{5}{*}{\begin{sideways}  $3$
and $4$\end{sideways}} & $0.1\%$ & {\it 18}  & 12     & 15 & 32   \\
& & $1\%$ & 15 & {\it 17}  & 12 & 19  \\
& & $5\%$ & 11 & 15  & {\it 16} & 11  \\
& & $30\%$ & 28  & 18  & 12 & {\it 16}  \\
\hline\hline
\end{tabular}
\caption{Break-even correlations of the First 2nd-To-Default in the
four-name basket ($\bar\sigma=50\%$). \label{F2TDTable}}
\end{center}
\end{table}

\begin{table}
\begin{center}
\begin{tabular}{||lll||c|c|c|c||}
\hline\hline \multicolumn{7}{||r||}{Spot Default intensities of
$1$ and $2$} \\ \hline & & & $0.1\%$ & $1\%$ & $5\%$ & $30\%$
\\ \hline
\multirow{5}{*}{\begin{sideways} Intensities \end{sideways}} &
\multirow{5}{*}{\begin{sideways}  $3$
and $4$\end{sideways}} & $0.1\%$ & {\it 20} & 22  & 51 & {\bf 70}  \\
& & $1\%$ & 17 & {\it 18}  & 25  & 69  \\
& &$5\%$ & 16 & 15  & {\it 16} & 53  \\
&  & $30\%$ & {\bf 29} & 25  & 19 & {\it 18}  \\
\hline\hline
\end{tabular}
\caption{Break-even correlations of the First 3rd-To-Default in the
four-name basket ($\bar\sigma=50\%$). \label{F3TDTable}}
\end{center}
\end{table}

\medskip

Moreover, we have slightly modified the previous empirical
framework: now, the basket has got ten names, and we have considered
a particular trajectory of individual survival probabilities
under~(\ref{BEDynamics}) (same time schedule). We have worked with
constant volatilities $\bar\sigma_t := \bar\sigma$ for all $t$. We
have assumed several scenarii in terms of individual spread
volatilities, spread correlations $\rho_{ij}$ and spot intensity
levels. Depending on the homogeneity among the ten underlying names,
we get different shapes in terms of break-even correlations (or
equivalently, in terms of beta factors).

\medskip

\begin{figure}
\begin{center}
\includegraphics[width=16cm,height=18cm]{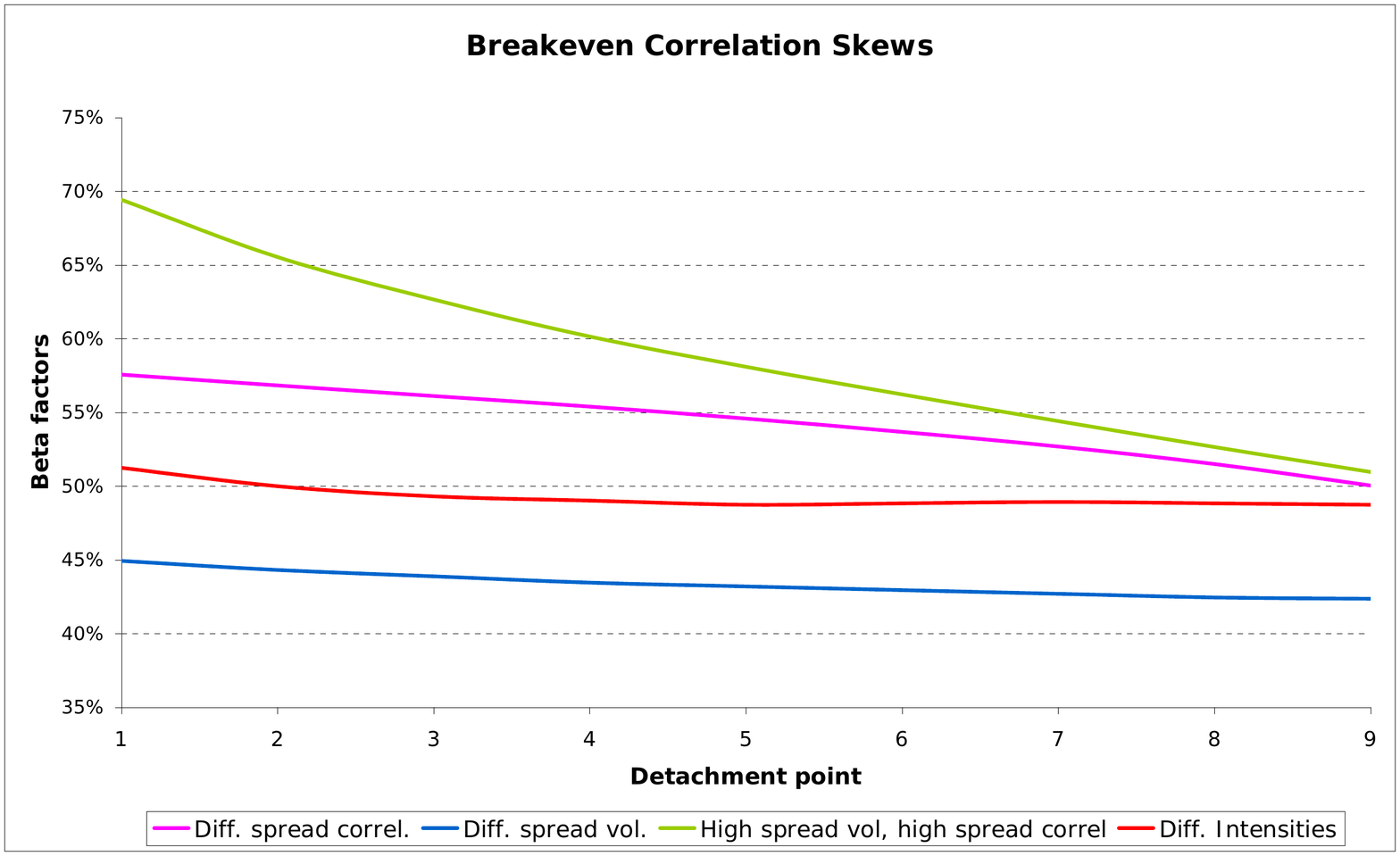}
\caption{Break-even "beta factors" on a simulated trajectory \label{BESkews0}}
\end{center}
\end{figure}

\begin{figure}
\begin{center}
\includegraphics[width=16cm,height=18cm]{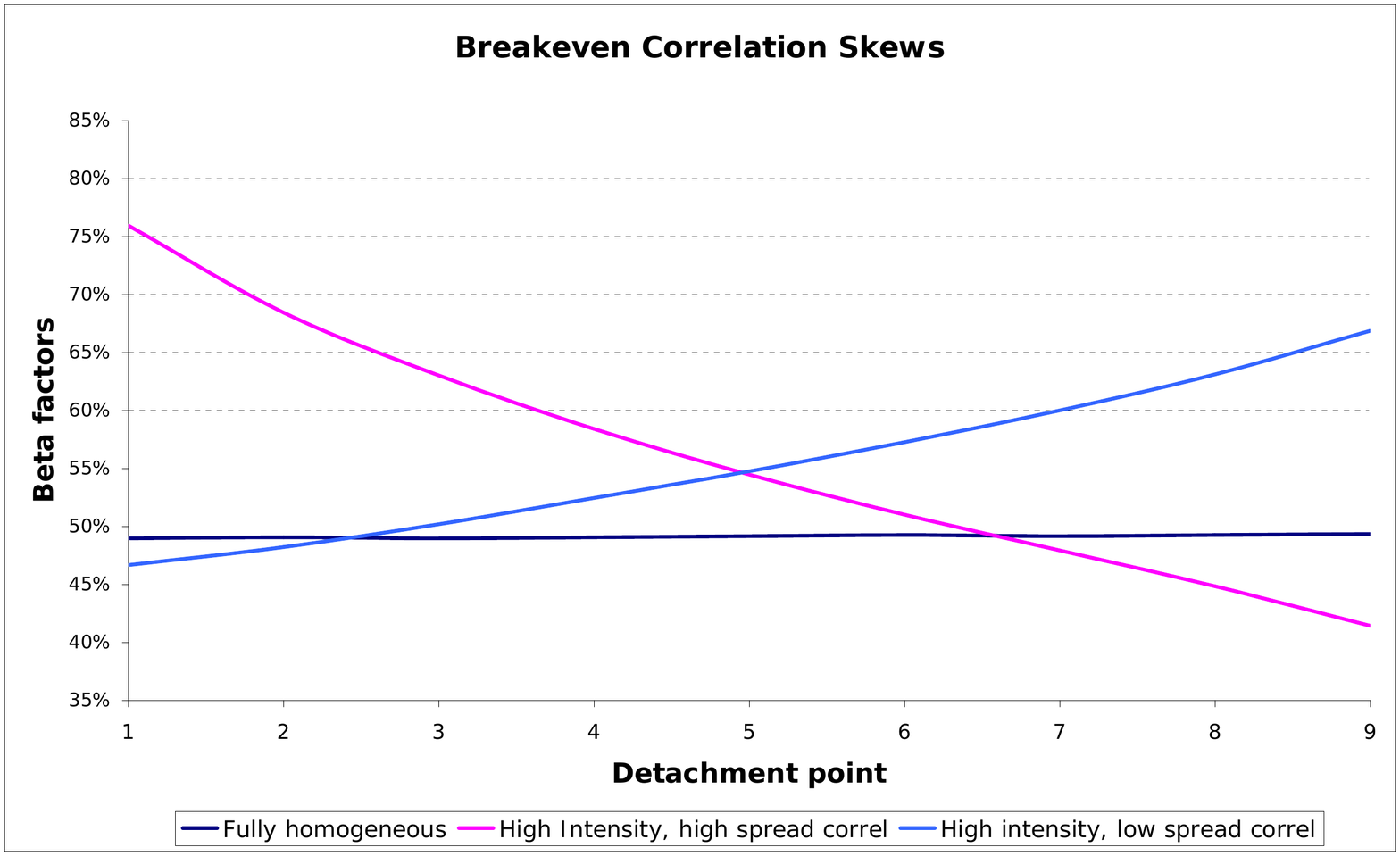}
\caption{Break-even "beta factors" on a simulated trajectory \label{BESkews}}
\end{center}
\end{figure}

\begin{figure}
\begin{center}
\includegraphics[width=16cm,height=18cm]{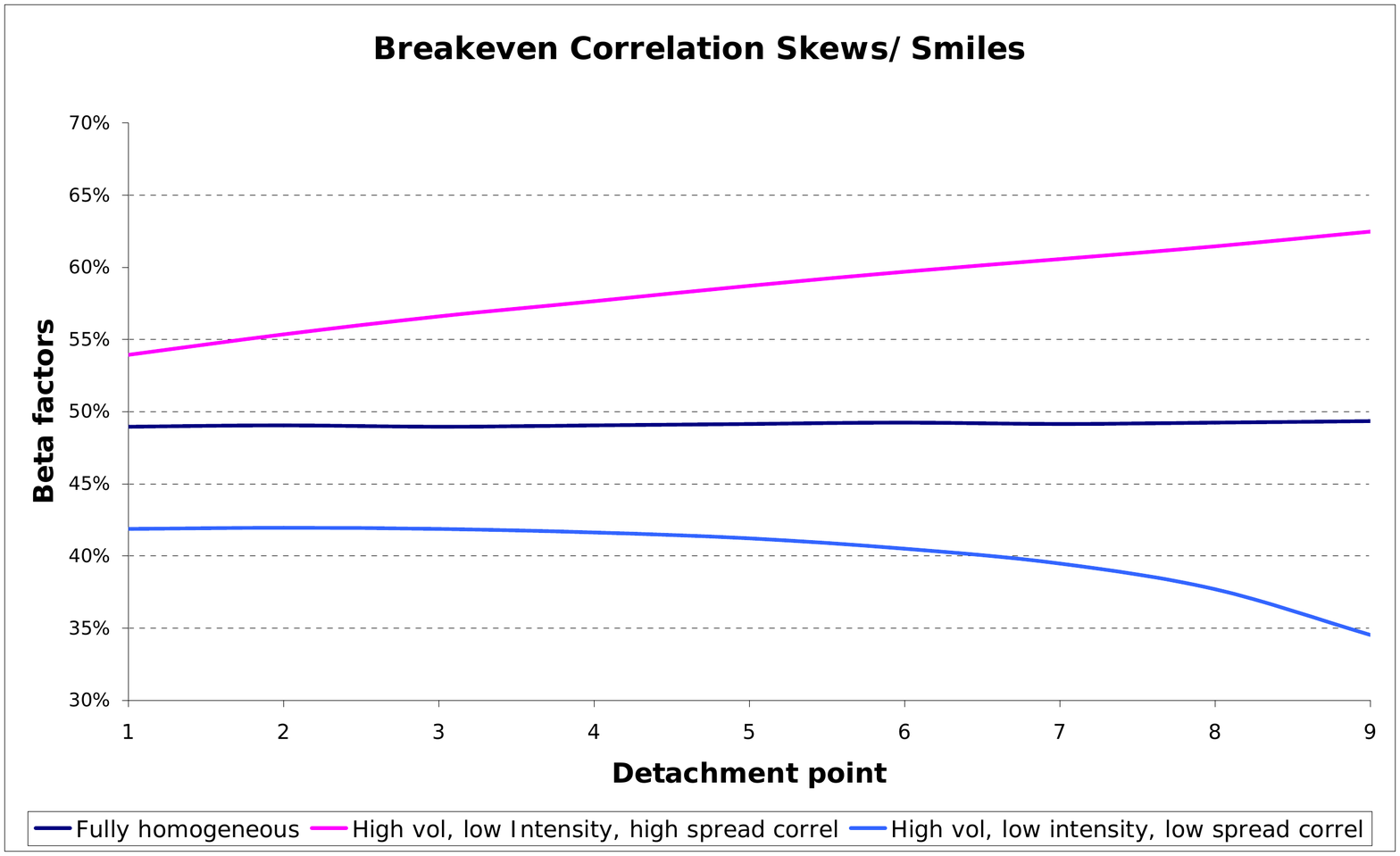}
\caption{Break-even "beta factors" on a simulated trajectory \label{BESkews2}}
\end{center}
\end{figure}

The reference case is related to a fully homogeneous basket, with
default intensities of $\lambda_0=5\%$, constant volatilities of
$\bar\sigma=50\%$ and constant spread correlations of
$\rho_{ij}=50\%$ for all $i\neq j$. In this case, the "break-even
beta factor"~\footnote{i.e. the squared root of the break-even
correlation} is closed to $50\%$ and independent from the strike, as
expected. Then, to move from this core scenario away, we have
applied some deformations on these numbers: see
tables~\ref{ScenarioVol},~\ref{ScenarioIntens}
and~\ref{ScenarioCorrel}. Broadly speaking, these scenarii apply
some rescaling method on all the parameters above, but differently
from name to name. We observe different break-even correlation skews
and smiles, depending on the heterogeneity of the underlying basket.
For instance, in figure~\ref{BESkews0}, different spread
volatilities, spread correlations or initial default intensities per
name (all other things being equal) generate downward sloping
break-even correlations. This effect is more striking when high
spread volatilities are associated with high spread correlations.
Nonetheless, we can produce upward sloping break-even correlation
skews by putting low beta factors on names that have got high
intensities: figure~\ref{BESkews}. Note we get the opposite when the
high beta factors are associated with high intensities. By combining
these three dimensions, we can even generate some unexpected
smiles/skews that are difficult to be forecasted:
figure~\ref{BESkews2}.

\begin{table}
\begin{center}
\begin{tabular}{||l||c|c|c||}
\hline\hline    Name & Core & Up & Down \\ \hline\hline
 1 & 0.50&  0.22&   1.13 \\
2& 0.50&    0.22&   1.13    \\
3 & 0.50&   0.33&   0.75    \\
4 & 0.50&   0.33&   0.75    \\
5 & 0.50&    0.50&   0.50    \\
6 & 0.50&   0.50&   0.50    \\
7 & 0.50&   0.75&   0.33    \\
8 & 0.50&   0.75&   0.33    \\
9 & 0.50 &  1.13 &0.22  \\
10 & 0.50 & 1.13    & 0.22\\
\hline\hline
\end{tabular}
\caption{Scenarii in terms of spread volatilities.
\label{ScenarioVol}}
\end{center}
\end{table}

\begin{table}
\begin{center}
\begin{tabular}{||l||c|c|c||}
\hline\hline    Name & Core & Up & Down \\ \hline\hline
1 & 0.05&   0.02&   0.11    \\
2&0.05& 0.02&   0.11    \\
3&0.05& 0.03&   0.08    \\
4& 0.05&    0.03&   0.08    \\
5& 0.05&    0.05&   0.05    \\
6& 0.05&    0.05&   0.05    \\
7& 0.05&    0.08&   0.03    \\
8& 0.05&    0.08&   0.03    \\
9& 0.05&    0.11&   0.02    \\
10& 0.05&   0.11&   0.02    \\
\hline\hline
\end{tabular}
\caption{Scenarii in terms of default intensities.
\label{ScenarioIntens}}
\end{center}
\end{table}

\begin{table}
\begin{center}
\begin{tabular}{||l||c|c|c||}
\hline\hline    Name & Core & Up & Down \\ \hline\hline
1& 0.50&    0.22&   0.99    \\
2& 0.50 & 0.22& 0.99    \\
3& 0.50&    0.33&   0.75    \\
4& 0.50&    0.33&   0.75    \\
5& 0.50&    0.50&   0.50    \\
6&0.50 &    0.50&   0.50\\
7& 0.50&    0.75&   0.33\\
8& 0.50&    0.75&   0.33\\
9& 0.50&    0.99 &  0.22\\
10& 0.50&   0.99 &  0.22\\
\hline\hline
\end{tabular}
\caption{Scenarii in terms of beta factors. \label{ScenarioCorrel}}
\end{center}
\end{table}

\medskip

We have calculated historical series of Break-even correlations, on
two standard credit indices: ITraxx S8 Master, and CDX IG9 Main,
both between their launch in September 2007 and June 2009. In
practice, we have calculated weekly P\&Ls that are generated by
continuously delta-hedging a base tranche $[0,x\%]$ (for ITraxx) or
a senior tranche $[x\%,100\%]$ (for CDX). The hedging instruments
are usual CDS on every name of the basket, and every relevant
maturity. The CDO tranche calculation is led for a grid of constant
correlation levels, from $0\%$ to $95\%$. We calculate P\&L
increments with a rolling window of six weeks. For every window, the
(empirical) break-even correlation is the level of flat correlation
that allows a zero P\&L variation~\footnote{Actually, to get nicer
results visually, we have smoothed break-even correlations: at every
date, the mapped break-even correlation is the weighted average of
the current and three previous ones. The weights are exponential
$\exp(-\lambda k)$, $k=0,1,2,3$ and $\lambda=0.3$.}. Moreover, to
simplify, we assume that no interest are generated in the cash
account, and that the running premia of all the instruments (CDS and
tranches) are zero. Thus, they are managed fully with upfront
paiements.

\medskip

\begin{figure}
\begin{center}
\includegraphics[width=16cm,height=18cm]{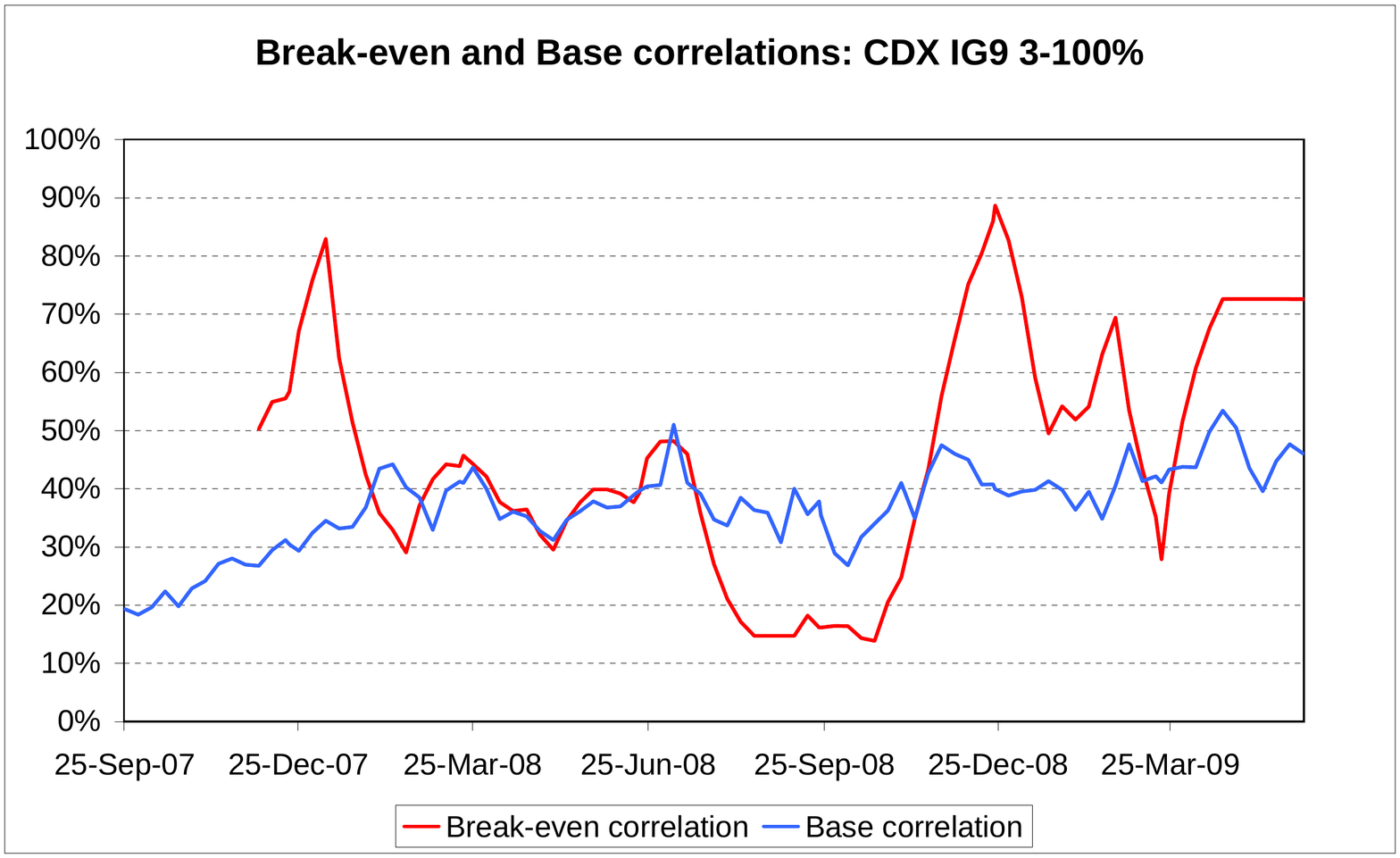}
\caption{Break-even correlations and base correlations, CDX IG9,
$3\%-100\%$. \label{CDXIG9_3p}}
\end{center}
\end{figure}

\begin{figure}
\begin{center}
\includegraphics[width=16cm,height=18cm]{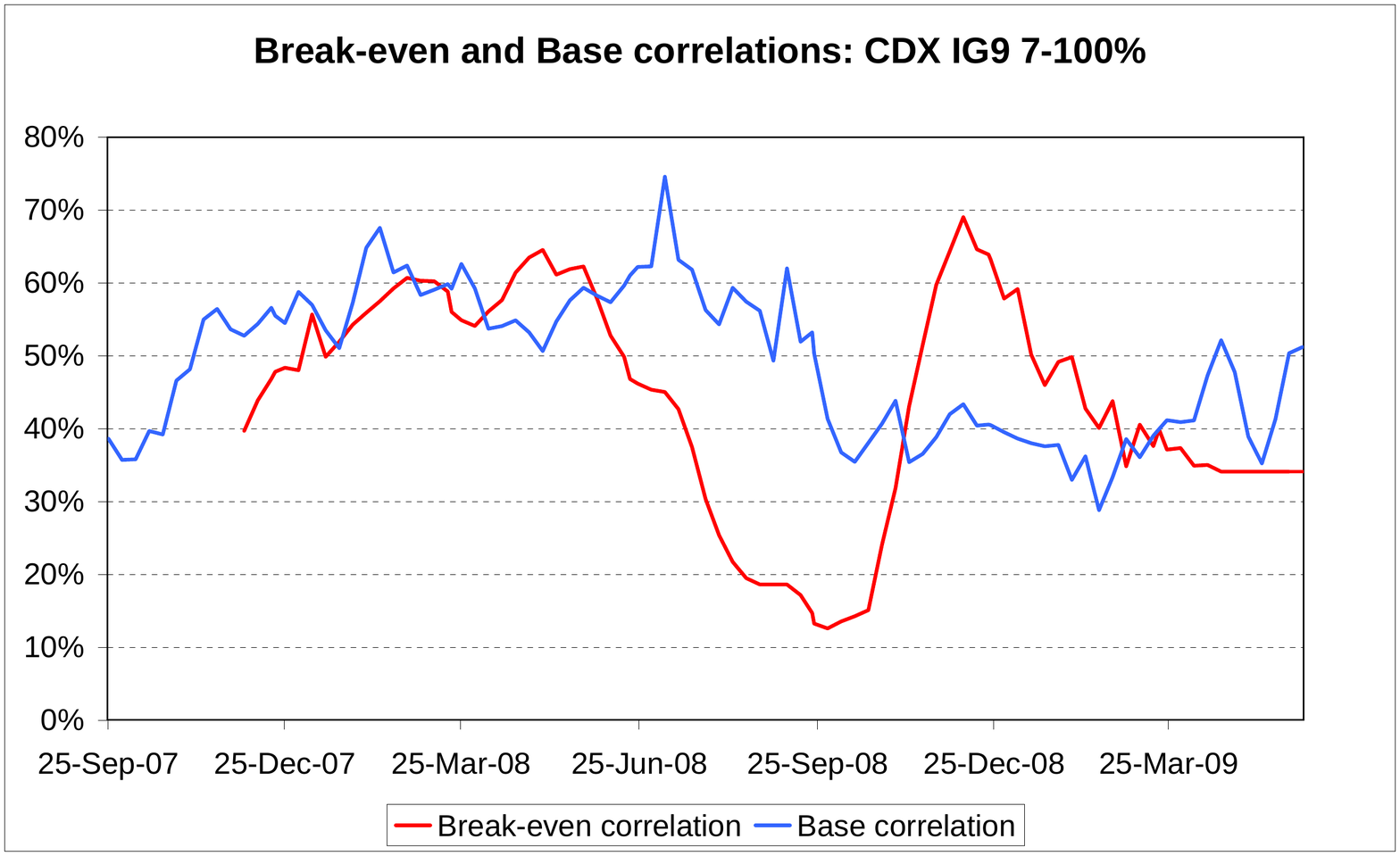}
\caption{Break-even correlations and base correlations, CDX IG9,
$7\%-100\%$. \label{CDXIG9_7p}}
\end{center}
\end{figure}

\begin{figure}
\begin{center}
\includegraphics[width=16cm,height=18cm]{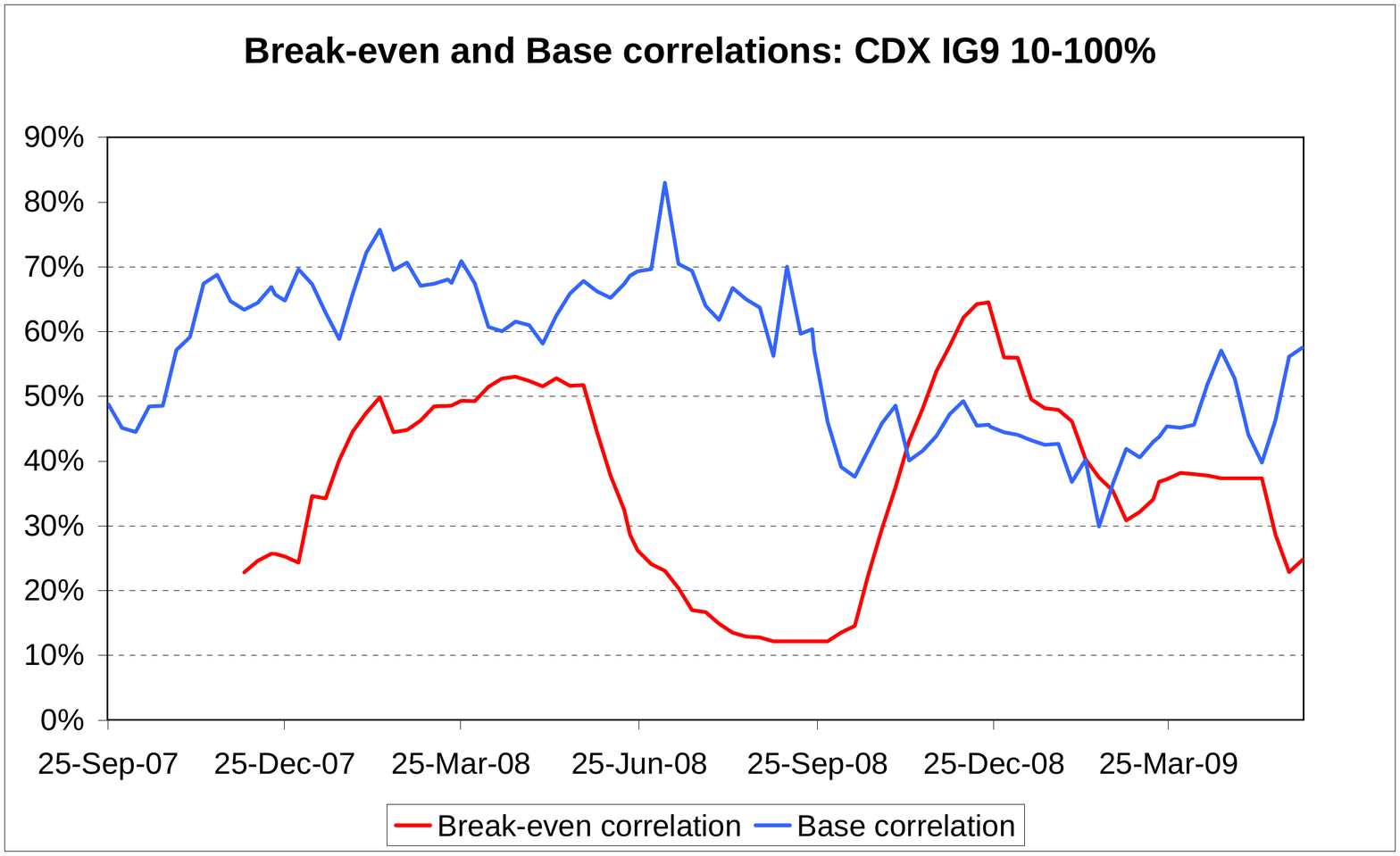}
\caption{Break-even correlations and base correlations, CDX IG9,
$10\%-100\%$. \label{CDXIG9_10p}}
\end{center}
\end{figure}

\begin{figure}
\begin{center}
\includegraphics[width=16cm,height=18cm]{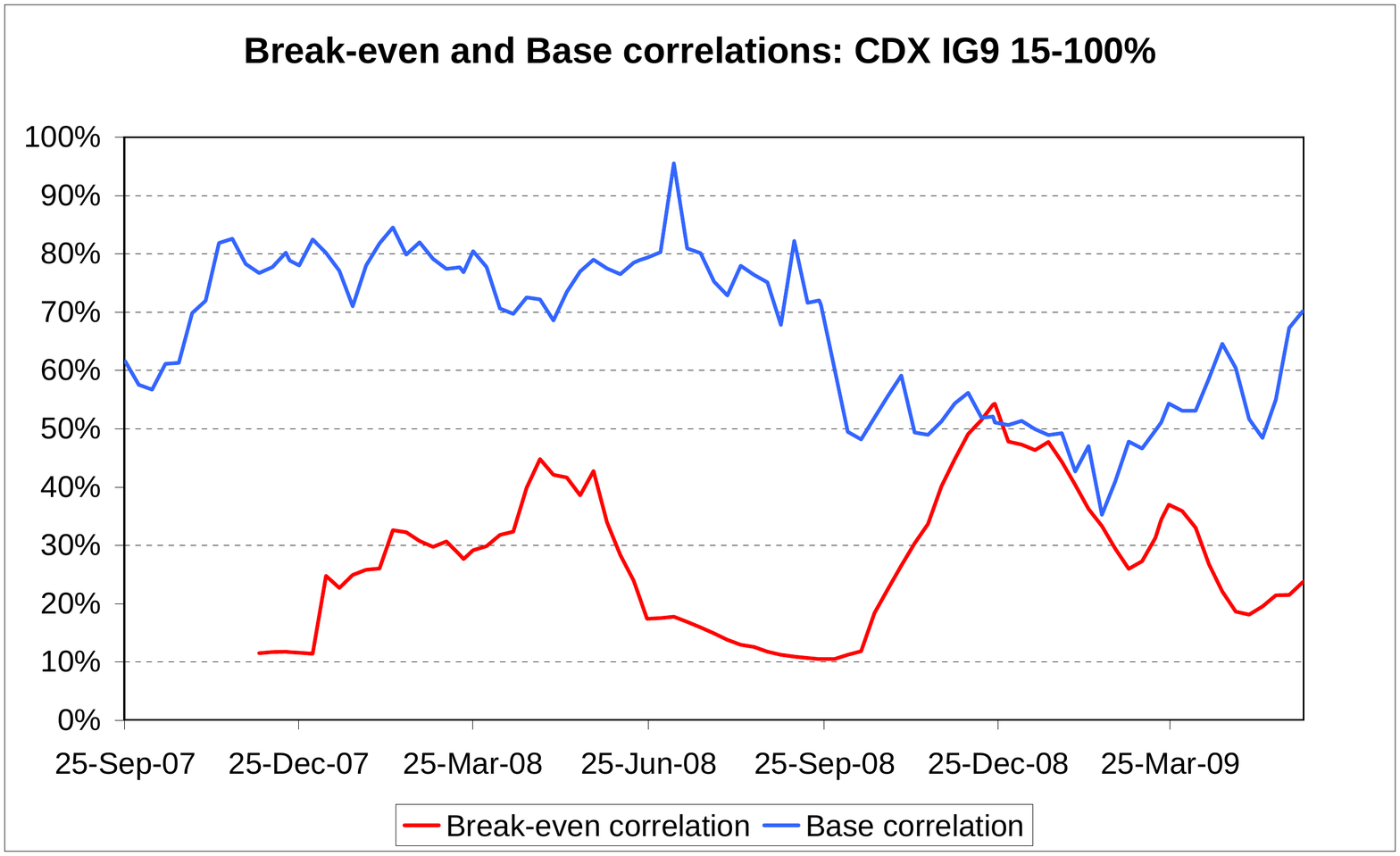}
\caption{Break-even correlations and base correlations, CDX IG9,
$15\%-100\%$. \label{CDXIG9_15p}}
\end{center}
\end{figure}

\begin{figure}
\begin{center}
\includegraphics[width=16cm,height=18cm]{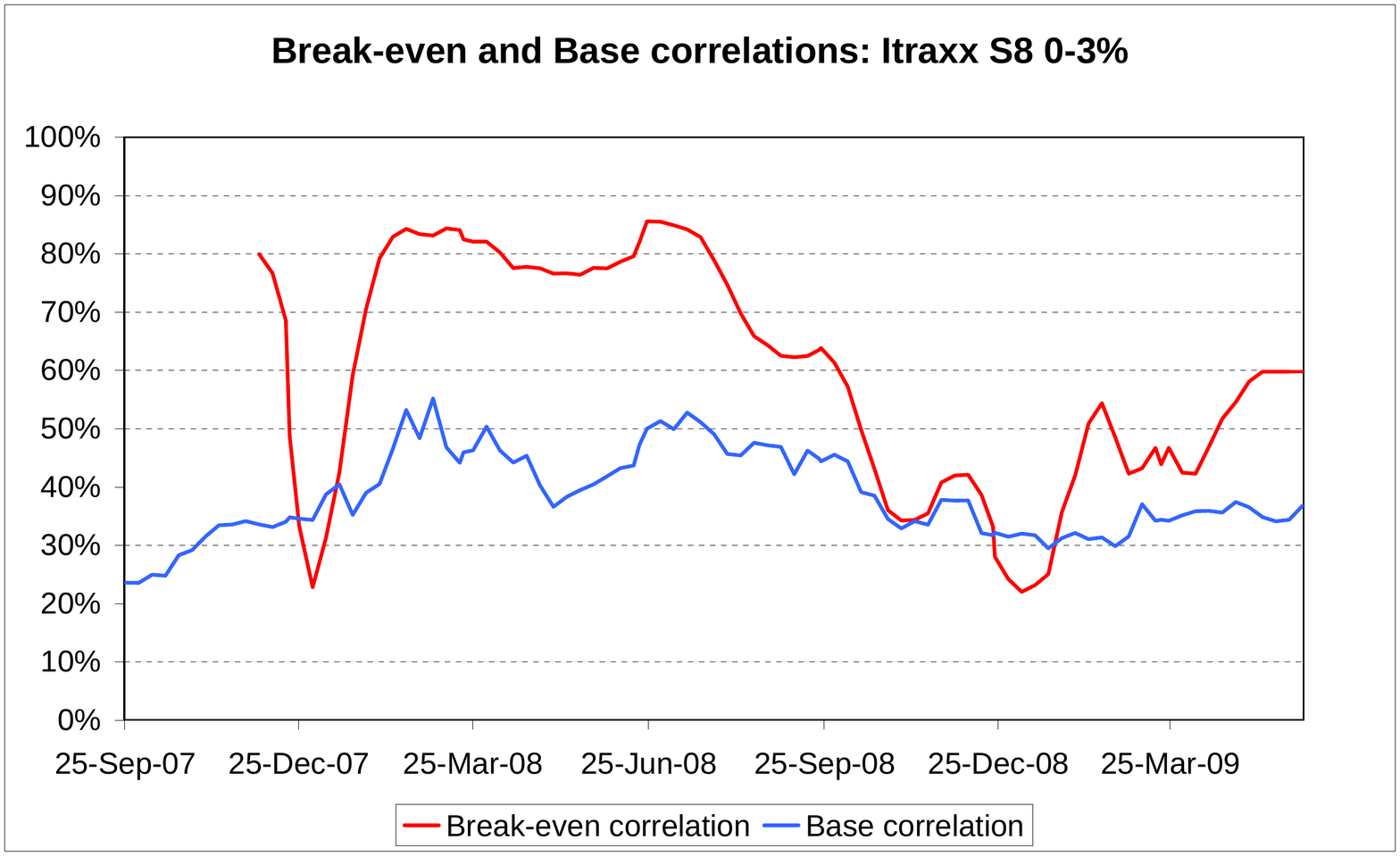}
\caption{Break-even correlations and base correlations, ITraxx
Main S8, $0-3\%$. \label{ITraxxS8_3p}}
\end{center}
\end{figure}

\begin{figure}
\begin{center}
\includegraphics[width=16cm,height=18cm]{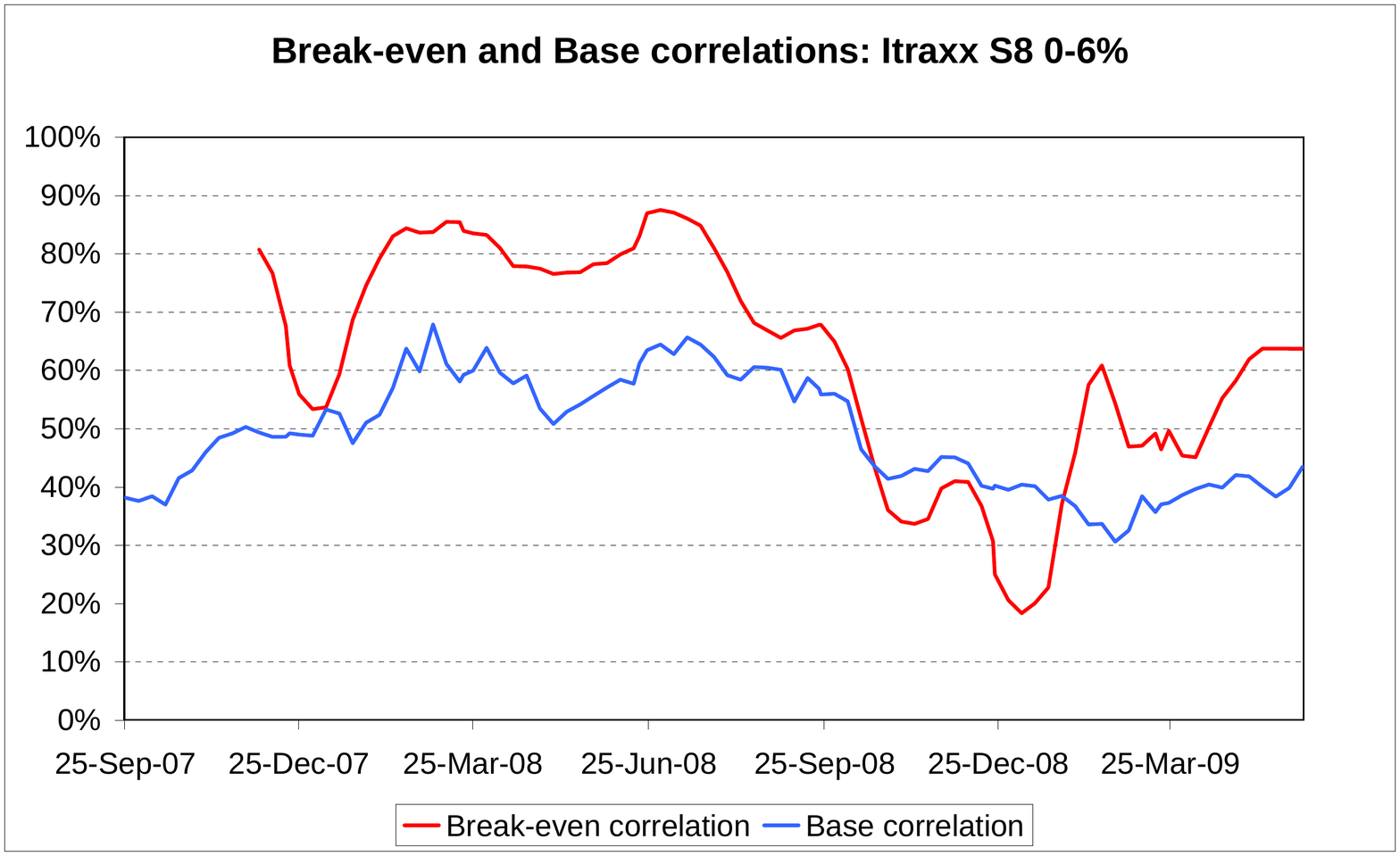}
\caption{Break-even correlations and base correlations, ITraxx
Main S8, $0-6\%$. \label{ITraxxS8_6p}}
\end{center}
\end{figure}

\begin{figure}
\begin{center}
\includegraphics[width=16cm,height=18cm]{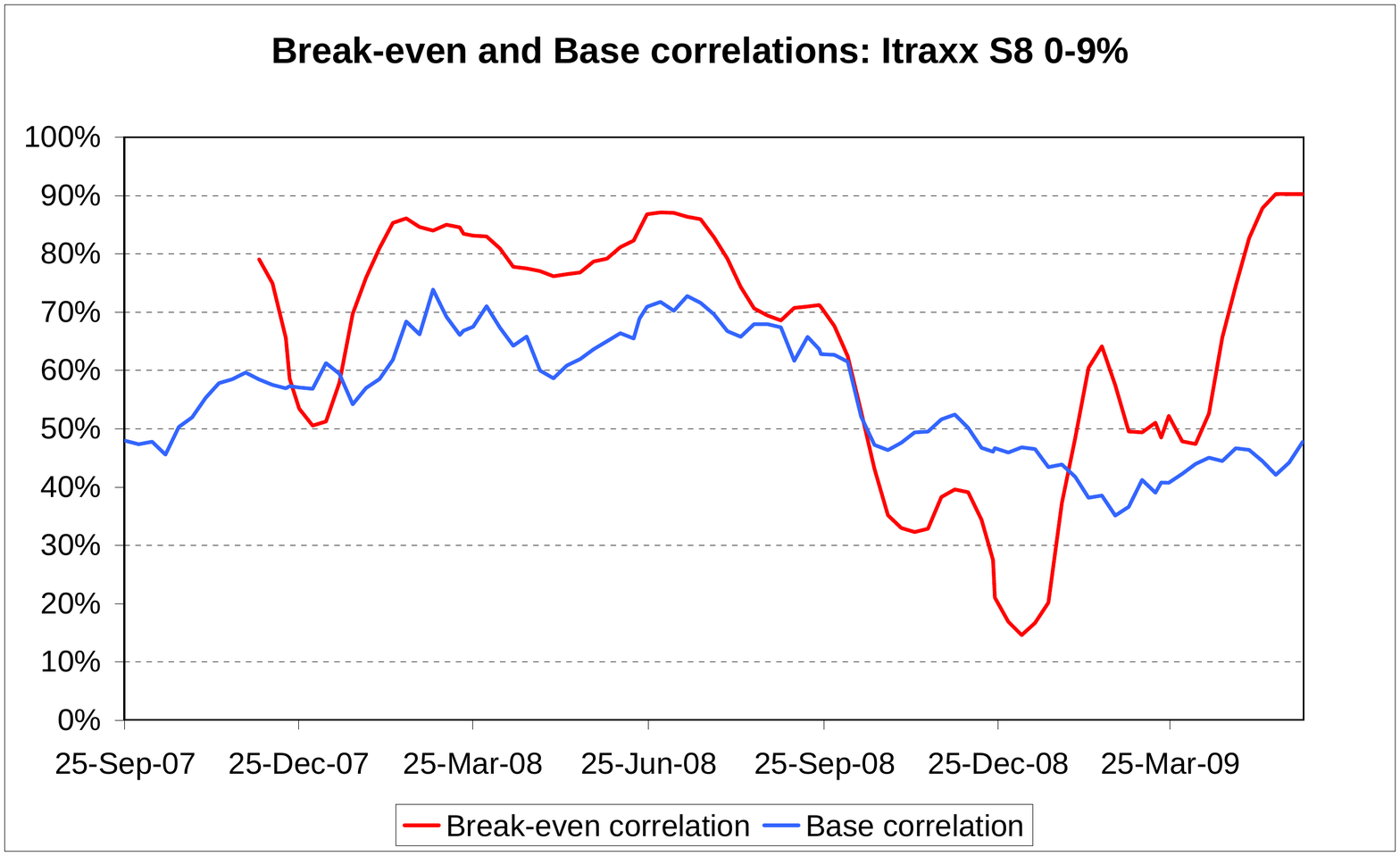}
\caption{Break-even correlations and base correlations, ITraxx
Main S8, $0-9\%$. \label{ITraxxS8_9p}}
\end{center}
\end{figure}

\begin{figure}
\begin{center}
\includegraphics[width=16cm,height=18cm]{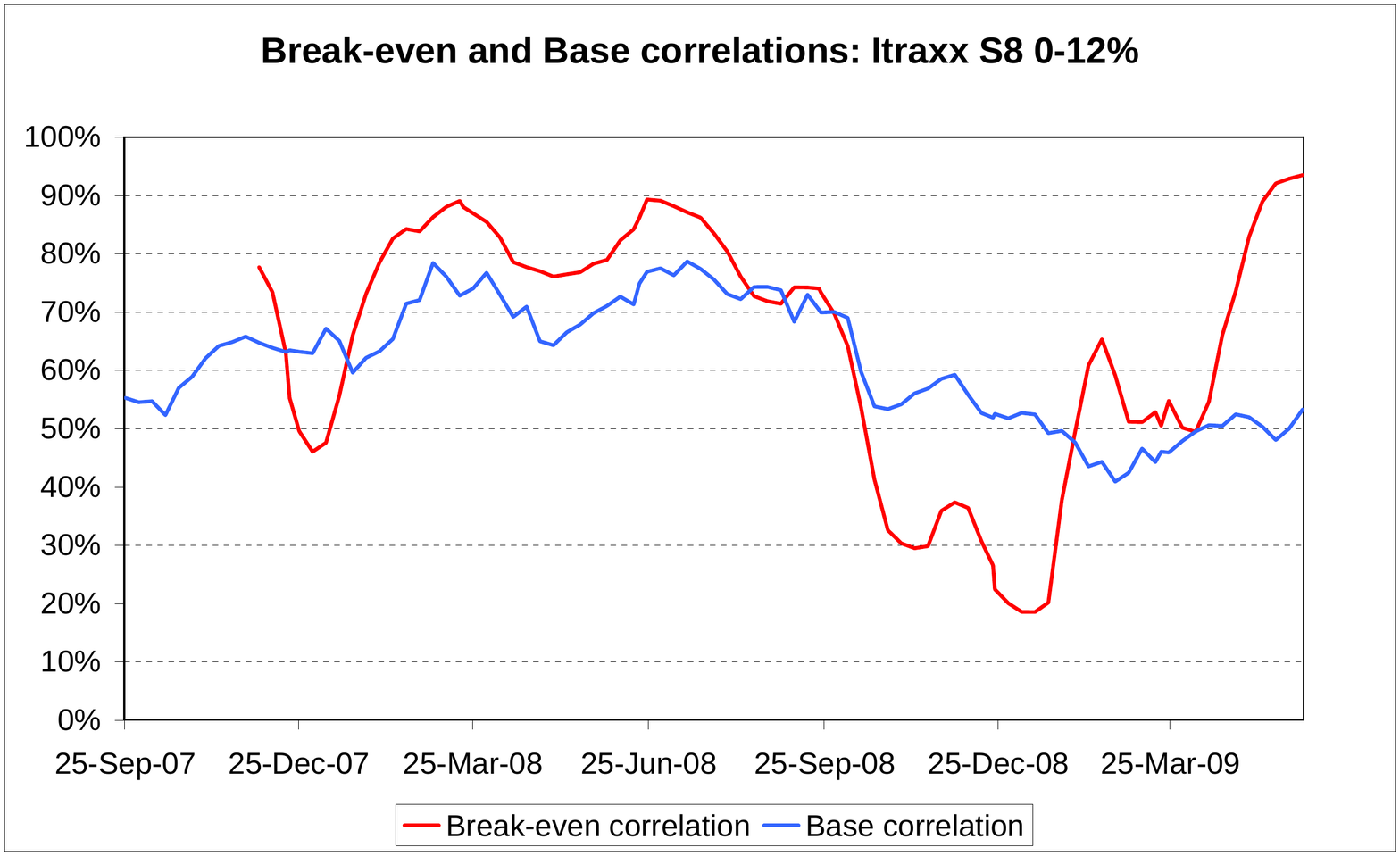}
\caption{Break-even correlations and base correlations, ITraxx
Main S8, $0-12\%$. \label{ITraxxS8_12p}}
\end{center}
\end{figure}

\begin{figure}
\begin{center}
\includegraphics[width=16cm,height=18cm]{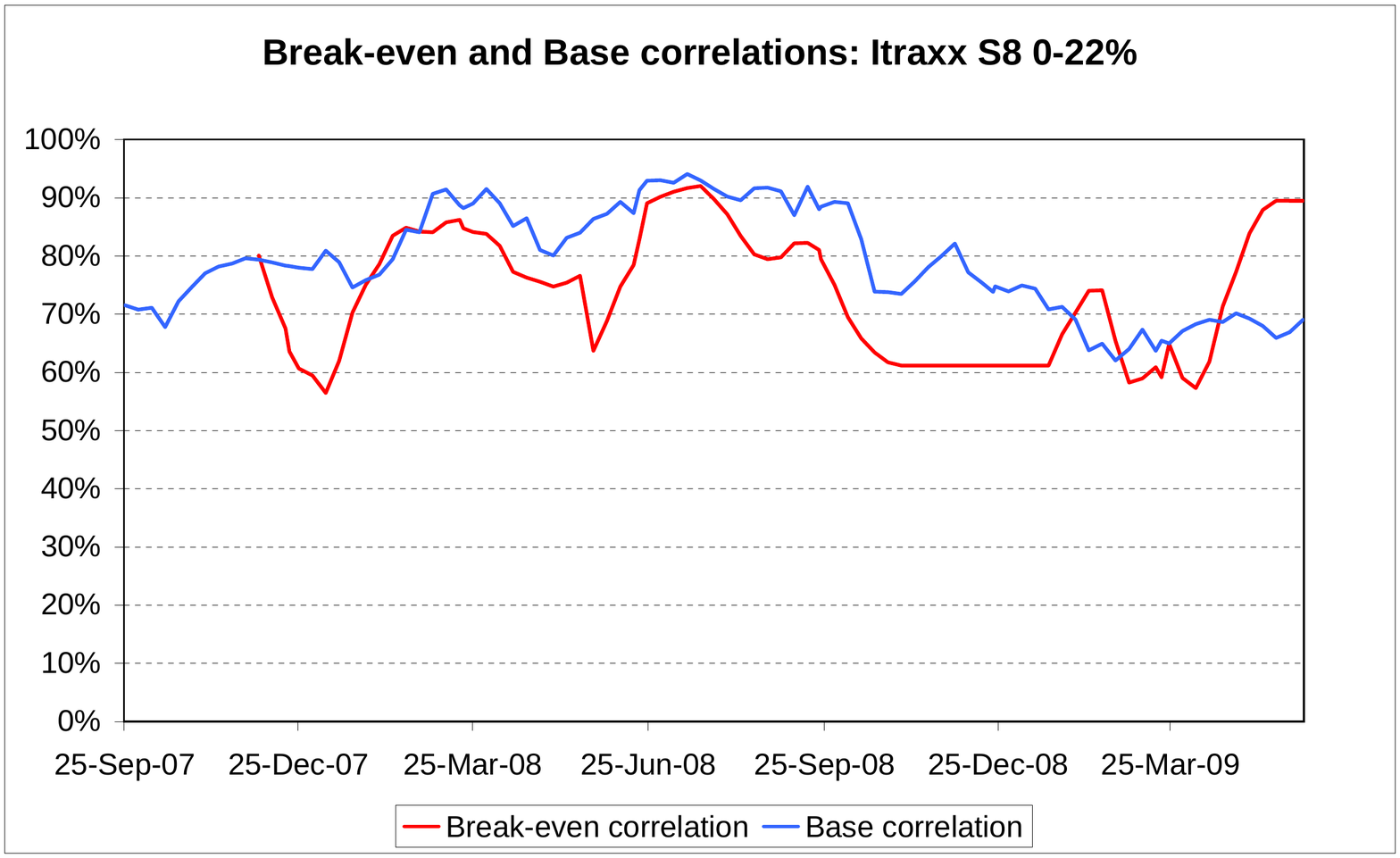}
\caption{Break-even correlations and base correlations, ITraxx
Main S8, $0-22\%$. \label{ITraxxS8_22p}}
\end{center}
\end{figure}

For the sake of comparison, on our graphs, we have shown the
break-even correlations and the associated base correlations, as
deduced from the market: see
figures~\ref{CDXIG9_3p},~\ref{CDXIG9_7p},~\ref{CDXIG9_10p},~\ref{CDXIG9_15p}
for the CDX IG9 (main),
and~\ref{ITraxxS8_3p},~\ref{ITraxxS8_6p},~\ref{ITraxxS8_9p},~\ref{ITraxxS8_12p},~\ref{ITraxxS8_22p}
for the index ITraxx S8 (main). It can be observed that the two
series move consistently, even if the break-even series generate
more volatile moves. But in average, the market base correlations
are not so different from break-even ones, as expected, except for
CDX senior tranches. It may be possible that investors in the latter
tranches require an additional risk premium, or, in other words,
overestimate systemic risk.

\medskip

In this practical experiment, several factors move us from the
theoretical framework away: the CDO tranches are not
European~\footnote{thus, pricing formulas include immediate payments
when default events occur}, heterogeneity in terms of recovery rates
for CDX, some default events have been observed during this period
of time for CDX~\footnote{Fannie Mae, Freddy Mac, Washington
Mutual}, and have not been fully anticipated by the market.
Moreover, our base correlations depend on the BNP-Paribas current
analytics that have evolved in this period of
time~\footnote{particularly, a switch towards a stochastic recovery
model in the last quarter of $2008$}. And, at the end of 2007, it
has sometimes been difficult to find implied levels of base
correlations from market quotes. Nonetheless, especially for ITraxx,
it appears that the concept of Break-even correlation can be a very
useful relative value tool. Indeed, theoretically, its moves should
be in line with those of the market base correlations. It is what we
observe empirically, broadly speaking. Thus, significant and
persistent gaps between both types of correlations should help to
forecast future moves of base correlations, because both series
should converge towards each other finally.

\section{Conclusion}
We have shown under which conditions usual ``static'' credit
portfolio models can be seen as replication models. Explicit
expressions of dynamic hedging errors have been provided for a large
class of portfolio derivatives. In particular, we have investigated
the one factor Gaussian copula model, the market standard, and its
multivariate extension. We have exhibited the unique family of
dynamics (in terms of survival probabilities), that are consistent
with a pricing by replication. We have shown that the matrix of
spread correlations $\rho_{ij}$ and some spread ``volatility-type''
coefficients $\bar\sigma_i$ should be combined in a particular way
to get break-even correlations. Broadly speaking, the matrix
$[2\bar\sigma_i \bar\sigma_j \rho_{ij}/(\bar\sigma_i^2 +
\bar\sigma_j^2)]$ should be definite positive. When it is the case,
this matrix is our break-even correlation matrix and should be used
for pricing and hedging purpose. We have shown that the one-factor
copula is more tractable than most other factor models and why it is
so difficult to find an alternative to it~\footnote{at least inside
the family of static models} by proving that most other models will
not lend themselves to replication.

\medskip

There remains a lot of avenues for further research. In particular,
we should try to adapt our framework to deal with
\begin{enumerate}
\item[(i)] richer dynamics in terms of survival probabilities, for instance by assuming underlying Levy processes,
\item[(ii)] random recoveries (Amraoui and Hitier 2008, Krekel 2008),
\item[(iii)] some extensions of the one-factor Gaussian copula framework, as random factor loadings (Andersen and Sidenius 2004) or local correlation (Turc et al. 2005),
\item[(iv)] perfect hedging when sudden jumps-to-default may occur before maturity.
\item[(v)] joint specification of all the $Q_{it}(T)$-dynamics for different maturities $T$, i.e. an arbitrage-free specification of term structures.
\end{enumerate}
All these points induce significant technical difficulties.
Concerning the point (i), additional jumps components would
challenge the desired perfect replication property. Moreover, random
recoveries cannot be tackled easily. At first glance, it could be
done by working with individual loss dynamics instead of survival
probabilities $Q_{it}(T)$. In point (iii), the idea is to build a
dependence between the pricing correlation level and the common
factor, i.e. to price with some functions $\rho_i(x)$,
$i=1,\ldots,n$. Point (iv) has been discussed in
section~\ref{General}. Since it should involve alternative
sensitivities and alternative instruments, this task has been left
for future research. The last point (v) is probably the most crucial
one. All our analysis has been led with a given maturity $T$. With
two maturities $T<T'$ and the same framework, we can check that it
is not possible to satisfy the arbitrage relation $Q_{it}(T) \geq
Q_{it}(T')$, for all times $t<T$. Indeed, when $t$ tends to $T$,
$Q_{it}(T)$ will tend to zero with a certain probability, when
$Q_{it}(T')$ will not exhibit such a feature (as a consequence of the
spread dynamics we have found). To find other model specifications, similar to first-hitting time models for instance, is the main challenge to tackle the multi-period formulation of the problem.

\medskip

\begin{center}
The authors thank their colleagues and former colleagues of BNP-Paribas and JP-Morgan for their help and fruitful discussions.
Bruno Bouchard, St\'ephane Cr\'epey, Peter J\"{a}ckel, Monique Jeanbleanc, Jean-Paul Laurent, Dong Li and
Marek Rutkowski have provided us highly valuable comments too.
\end{center}

\appendix

\section{Proof of theorem~\ref{thdpit2}}
\label{thgeneral}

By the conditional independence property, the price of the structured product is
$$ V_t  = \sum_{\bdelta(T)} \psi(\bdelta (T) ) \int \Prod_{j=1}^n  p_{j|x}^{\delta_j}q_{j|x}^{1-\delta_j} f_X(x) \, dx  =V(Q_1,\ldots,Q_n),$$
with our notations. We have
\begin{align}
\frac{\partial^2 V_t}{\partial^2 Q_i} &=
 \sum_{\bdelta(T)} \psi(\bdelta (T) ) \int \Prod_{j\neq i}
 p_{j|x}^{\delta_j}q_{j|x}^{1-\delta_j}.(2 \delta_i-1)
\frac{\partial^2 p_{i|x}}{\partial^2 Q_i} f_X(x) \, dx ,
\end{align}
and, for every couple $(i,j)$, $i\neq j$,
\begin{align}
\frac{\partial^2 V_t}{\partial Q_i\partial Q_j} &=
 \sum_{\bdelta(T)} \psi(\bdelta (T) ) \int \Prod_{k\neq i,j}
 p_{k|x}^{\delta_k}q_{k|x}^{1-\delta_k}.(2 \delta_i-1)(2 \delta_j-1)
\frac{\partial p_{i|x}}{\partial Q_i}\cdot \frac{\partial p_{j|x}}{\partial Q_j} f_X(x) \, dx .
\end{align}
Then, by applying Ito's formula, we get
\begin{align}
 dV_t &=(\ldots)\cdot d\vec{W}_t + \frac{1}{2}\sum_{i,j}\frac{\partial^2 V_t}{\partial Q_i\partial Q_j}\sigma_{it} \sigma_{jt} \rho_{ij} dt  \nonumber\\
 &= (\ldots)\cdot d\vec{W}_t + \sum_{\bdelta(T)} \psi(\bdelta (T) ) \sum_{i=1}^n dt\int \Prod_{j\neq i}
 p_{j|x}^{\delta_j}q_{j|x}^{1-\delta_j}.(2 \delta_i-1)
 \frac{\sigma_i^2 }{2}\cdot
 \frac{\partial^2 p_{i|x}}{\partial^2 Q_i}  f_X(x) \, dx \nonumber\\
 &+  \frac{dt}{2}\sum_{i \neq j} \sum_{\bdelta(T)} \psi(\bdelta (T) ) \int \Prod_{k\neq i,j}
 p_{k|x}^{\delta_k}q_{k|x}^{1-\delta_k}.(2 \delta_i-1)(2 \delta_j-1) \sigma_i \sigma_j \rho_{ij}
 \frac{\partial p_{i|x}}{\partial Q_i}\cdot \frac{\partial p_{j|x}}{\partial Q_j} f_X(x) \, dx
\nonumber  \\
& := (\ldots)\cdot d\vec{W}_t + d\pi_1 + d\pi_2.  \label{Mydpit}
\end{align}

By an integration by parts and with our notations, we get for every $i$
\begin{eqnarray*}
\lefteqn{
\int \Prod_{j\neq i}  p_{j|x}^{\delta_j}q_{j|x}^{1-\delta_j}
  \frac{\partial^2 p_{i|x}}{\partial^2 Q_i}  f_X(x) \, dx   }\\
&=& \left[ \Prod_{j\neq i}  p_{j|x}^{\delta_j}q_{j|x}^{1-\delta_j} \chi_i(x)  \right]_{-\infty}^{+\infty}
- \sum_{j,j\neq i} \int \Prod_{k\neq i,j}
 p_{k|x}^{\delta_k}q_{k|x}^{1-\delta_k}.(2 \delta_j-1)
 \frac{\partial p_{j|x}}{\partial x} \chi_i(x) \, dx .
\end{eqnarray*}
We deduce
\begin{align*}
d\pi_1 &= \eta -\frac{dt}{2}\sum_{\bdelta(T)} \psi(\bdelta (T) ) \sum_{i<j} \int \Prod_{k\neq i,j}
 p_{k|x}^{\delta_k}q_{k|x}^{1-\delta_k}.(2 \delta_i-1)(2 \delta_j-1)
 \left[ \sigma_i^2  \frac{\partial p_{j|x}}{\partial x} \chi_i(x) +
 \sigma_j^2 \frac{\partial p_{i|x}}{\partial x} \chi_j(x) \right]  \, dx ,
\end{align*}
and the formula follows. $\Box$

\medskip

\section{Proof of equation~(\ref{technicalCopGauss})}
\label{technical}
To lighten the notations, set $\bar{s}_j := \bar\Phi^{-1}(Q_{jt})$.
Simple calculations provide
$$  \frac{\partial p_{j|x}}{\partial Q_j}= - \frac{\phi(d_j)}{\phi(\bar s_j) \sqrt{1-\rho_j^2}},  $$
and
$$  \frac{\partial^2 p_{j|x}}{\partial^2 Q_j}=
\frac{\phi(d_j)}{\phi(\bar s_j)^2 (1-\rho_j^2)^{3/2}}\rho_j[ x - \rho_j \bar\Phi^{-1}(Q_j)].  $$
Thus,
$$ \chi_j(x) =
\frac{\rho_j}{\phi(\bar s_j)^2 (1-\rho_j^2)^{3/2}}\int_{-\infty}^x [ u - \rho_j \bar s_j].\phi(d_j)\phi(u) \, du  .$$
But a key point in favor of the Gaussian copula specification is that
\begin{equation}
\phi(d_j)\phi(x)=\phi\left(\frac{x-\rho_j \bar s_j }{\sqrt{1-\rho_j^2}} \right) \phi(\bar s_j).
\end{equation}
Therefore, we deduce
\begin{align*}
\chi_j(x) &=
\frac{\rho_j}{\phi(\bar s_j) (1-\rho_j^2)^{3/2}}\int_{-\infty}^x [ u - \rho_j \bar s_j]
.\phi\left(\frac{u-\rho_j \bar s_j }{\sqrt{1-\rho_j^2}} \right) \, du  \\
&= -\frac{\rho_j \phi(d_j) \phi(x)}{\phi(\bar s_j)^2 (1-\rho_j^2)^{1/2}} = \frac{\rho_j \phi(x)}{\phi(\bar s_j)}\cdot
\frac{\partial p_{j|x}}{\partial Q_j}\cdot\hspace{3cm}
\end{align*}
Moreover, we get easily
$$ \frac{\partial p_{j|x}}{\partial x} = \rho_j \phi(\Phi^{-1}(Q_{jt}))\frac{\partial p_{j|x}}{\partial Q_j},$$
and the result follows from theorem~\ref{thdpit2}.$\Box$

\medskip

\section{Proof of theorem~\ref{Unicity}.}
\label{UniqueDynamics}
Clearly, conditions $(i)$, $(ii)$ and $(iii)$ are sufficient so that $(V_t)$ is a $\F$-martingale under $Q$, invoking theorem~\ref{theocopG}.

\medskip

Now, let us prove the necessity of such conditions.
We want to cancel the drift of $dV_t$ for any random trajectory of all $(Q_{it})_{t\in I}$, $I\subset [0,T)$, $i=1,\ldots,n$.
To fix the ideas and without a lack of generality, we consider the particular
couple of indices $(1,2)$. In particular, the drift of $dV_t$ should be zero for trajectories where the $Q_{it}$ take very large or very small
values, $i\neq 1,2$. To be specific, we will consider trajectories such that $ \sup_x\phi(d_i)\phi(x) < \varepsilon$, for a given
$\varepsilon >0$, and when $i \neq 1,2$. Thus, since we are under the ``no default'' assumption, under our assumptions, the drift of $dV_t$ is
\begin{align*}
& \frac{dt}{2}\left[ 2 \beta_{1t} \beta_{2t} \rho_{12}
 -\rho_1 \rho_2 \beta_{1t}^2  - \rho_1\rho_2 \beta_{2t}^2  \right]\sum_{\bdelta(T)} \psi(\bdelta (T) )  (2 \delta_1-1)(2 \delta_2-1) \\
 &\cdot \int \Prod_{k\neq 1,2} p_{k|x}^{\delta_k}q_{k|x}^{1-\delta_k}
\frac{\phi(d_1)\phi(d_2)\phi(x)}{\sqrt{1-\rho_1^2}\sqrt{1-\rho_2^2}}\, dx   + O(\varepsilon).
\end{align*}
We have assumed that there exist some indicators $\delta_k^\ast$, $k\neq 1,2$ such that
$$ \psi(1,1,\bdelta^\ast_{-(1,2)}) - \psi(1,0,\bdelta^\ast_{-(1,2)}) - \psi(0,1,\bdelta^\ast_{-(1,2)})
+ \psi(0,0,\bdelta^\ast_{-(1,2)}) \neq 0.$$
Now, consider a compact subset $\Ac$ such that $ \int \1( x \not \in \Ac)
\phi(x) \, dx < \varepsilon$.
We can particularize our trajectories $Q_{it}$ even more, so that
\begin{itemize}
\item
$$ \sup_{x\in \Ac} \Prod_{k\neq 1,2}  p_{k|x}^{\delta_k}q_{k|x}^{1-\delta_k} \leq \varepsilon,$$
when $\delta_k \neq \delta_{k}^\ast$ for at least one index $k\neq 1,2$, and
\item
$$ \inf_{x\in \Ac} \Prod_{k\neq 1,2}  p_{k|x}^{\delta_k^\ast}q_{k|x}^{1-\delta_k^\ast} \geq 1/2.$$
\end{itemize}
This is possible, because, for every index $k\neq 1,2$, we can find $Q_{kt}$ such that $p_{k|x}$ (resp. $q_{k|x}$) is closed to one when $\delta_k^\ast=1$ ($\delta_k^\ast=0$), and uniformly w.r.t. $x$ in a compact subset.

\medskip

Therefore, the sum defining the hedging error is ``reduced'' even
more: for this particular choice of $Q_{kt}$, $k\neq i,j$, the drift
of $dV_t$ is
\begin{align*}
& \frac{dt}{2} \left( \psi(1,1,\bdelta^\ast_{-(1,2)}) - \psi(1,0,\bdelta^\ast_{-(1,2)})
-\psi(0,1,\bdelta^\ast_{-(1,2)}) + \psi(0,0,\bdelta^\ast_{-(1,2)})  \right) \\
&\cdot  \left[ 2 \beta_{1t} \beta_{2t} \rho_{12}
 -\rho_1 \rho_2 \beta_{1t}^2  - \rho_1 \rho_2 \beta_{2t}^2  \right] \int \Prod_{k\neq 1,2}
 p_{k|x}^{\delta^\ast_k}q_{k|x}^{1-\delta^\ast_k} \frac{ \phi(d_1)\phi(d_2)\phi(x)}{\sqrt{1-\rho_1^2}\sqrt{1-\rho_2^2}}\, dx   + O(\varepsilon) \\
&=  Cst \left[ 2 \beta_{1t} \beta_{2t} \rho_{12}
 -\rho_1 \rho_2 \beta_{1t}^2  - \rho_1 \rho_2 \beta_{2t}^2  \right] \int_{\Ac}
\phi(d_1)\phi(d_2)\phi(x)\, dx   + O(\varepsilon).
\end{align*}
Since $\varepsilon$ is arbitrarily small, this drift is zero if and only if
$$  2 \beta_{1t} \beta_{2t} \rho_{12} -\rho_1 \rho_2 [\beta_{1t}^2 +\beta_{2t}^2]  =0,$$
for every $Q_{1t}$ and $Q_{2t}$.

\medskip

Actually, it is possible to lead the same reasoning for all
couples $(i,j)$, $i\neq j$, under the assumption (P). Thus, we
recover equation~(\ref{GeneralCondition}), that is now a necessary
and sufficient condition:
\begin{equation}
2\beta_{it} \beta_{jt} \rho_{ij}  = [ \beta_{it}^2 + \beta_{jt}^2] \rho_i \rho_j ,
\label{EDPGCM2}
\end{equation}
for every $x$, $Q_{it}$, $Q_{jt}$ and every couple $(i,j)$, $i\neq j$.
The latter condition can be true if and only if the previous $\beta_{it}$ are independent of $Q_{it}$.
They can depend on time $t$, but iff $\beta_{it}/\beta_{jt}$ is independent of $t$ for all $(i,j)$, $i\neq j$.
Therefore, we get equation~(\ref{BEDynamics}), by rewriting $\beta_{it}$ as $\bar\sigma_{i}\xi_t$, for some positive function $\xi$
and some constant $\bar\sigma_{i}>0$.
Our assumption on $\Sigma^S$ provides the existence of the pricing parameters $\rho_i$, $i=1,\ldots,n$.

\medskip

To finish the proof, it remains to prove that $\lim_{t\rightarrow T}
\int_0^t \xi_{u}^2 \, du = +\infty$. By setting $Z_t = \Phi^{-1}(Q_{it})$ and invoking Ito's lemma, we see easily that
the solution of the PDE~(\ref{BEDynamics}) is given by
$$Q_{it} = \Phi\left(\exp(\frac{\bar\sigma_{i}^2 \int_0^t \xi_u^2 \, du}{2})
 \Phi^{-1}(Q_{i0}) +  \int_0^t \exp(\frac{\bar\sigma_{i}^2 \int_u^t \xi_s^2  \, ds}{2}) \bar\sigma_{i} \xi_u \, dW_{iu} \right),$$
 where $(W_{iu})$ is a Brownian motion under $Q$.
Note that $Q_{it}(T)$ tends to zero or one when $t$ tends to $T$. Assume that
$\lim_{t\rightarrow T} \int_0^t \xi_{u}^2 \, du = \ell
<+\infty$. Then, $ \int_0^T \exp(\bar\sigma_{i}^2\int_u^T \xi_s^2 \, ds/2)
\xi_u \, dW_{iu}$  is a non degenerate Gaussian r.v. and it
will be impossible to get $\lim_{t\rightarrow T} Q_{it} \in
\{0,1\}$ almost surely. So the result. $\Box$

\section{Proof of theorem~\ref{GentheocopG}.}
\label{Proof_GentheocopG}
We come back to the first steps of the proof of theorem~\ref{thdpit2} in section~\ref{thgeneral}.
With these previous notations, we have now
\begin{equation}
 \frac{d\pi_1}{dt} = \sum_{\bdelta(T)} \psi(\bdelta (T) ) \sum_{i=1}^n \int \Prod_{j\neq i}
 p_{j|\xx}^{\delta_j}q_{j|\xx}^{1-\delta_j}.(2 \delta_i-1)
 \frac{\sigma_i^2 }{2}\cdot
 \frac{\partial^2 p_{i|\xx}}{\partial^2 Q_i}  f_X(\xx) \, d\xx ,
 \end{equation}
and
\begin{equation}
\frac{d\pi_2}{dt} =  \frac{1}{2}\sum_{i \neq j} \sum_{\bdelta(T)} \psi(\bdelta (T) ) \int \Prod_{k\neq i,j}
 p_{k|\xx}^{\delta_k}q_{k|\xx}^{1-\delta_k}.(2 \delta_i-1)(2 \delta_j-1) \sigma_i \sigma_j \rho_{ij}
 \frac{\partial p_{i|\xx}}{\partial Q_i}\cdot \frac{\partial p_{j|\xx}}{\partial Q_j} f_X(\xx) \, d\xx.
\end{equation}
Under the $p$-factor Gaussian copula framework, these quantities can be rewritten:
\begin{equation}
\frac{d\pi_1}{dt} = \sum_{\bdelta(T)} \psi(\bdelta (T) ) \sum_{i=1}^n \int \Prod_{j\neq i}
 p_{j|\xx}^{\delta_j}q_{j|\xx}^{1-\delta_j}.(2 \delta_i-1)
 \frac{\sigma_i^2 }{2}\cdot
 \frac{\phi(d_i)[\rho_i'\xx - |\rho_i|^2\bar{s}_i]}{\phi(\bar{s}_i)^2 (1-|\rho_i|^2)^{3/2}}  \prod_{k=1}^p\phi(x_k) \, dx_k ,
 \end{equation}
where, as previously, we have set $\bar{s}_i := \bar\Phi^{-1}(Q_{it})$. Moreover,
\begin{eqnarray}
 \lefteqn{
\frac{d\pi_2}{dt} =  \frac{1}{2}\sum_{i \neq j} \sum_{\bdelta(T)} \psi(\bdelta (T) ) \int \Prod_{k\neq i,j}
 p_{k|\xx}^{\delta_k}q_{k|\xx}^{1-\delta_k}.(2 \delta_i-1)(2 \delta_j-1)  \nonumber }\\
 &\cdot & \sigma_i \sigma_j \rho_{ij}
 \frac{\phi(d_i)\phi(d_j)}{\phi(\bar{s}_i) \phi(\bar{s}_j) } \cdot
 \frac{\prod_{k=1}^p \phi(x_k) \, dx_k }{\sqrt{1-|\rho_i|^2}\sqrt{1-|\rho_j|^2}} \cdot
\label{Gendpi2}
\end{eqnarray}
The key point of the proof is to exhibit the right variables to make integration by parts.
Let us introduce some additional notations : for any vector $\xx$ in $\RR^p$ and $r\in \{1,\ldots,p\}$,
let us denote by $\xx_{-r}$ the $p-1$ vector $(x_1,\ldots,x_{r-1},x_{r+1},\ldots,x_p)$.
For every integer $r\leq p$, simple calculations provide the identity
\begin{equation}
\phi(d_i)\phi(x_r)= \phi\left(\sqrt{\frac{1-|\rho_{i,-r}|^2}{1-|\rho_{i}|^2}} \cdot
\{ x_r - \frac{\rho_{i,r}\bar{s}_{i,r}}{1-|\rho_{i,-r}|^2} \}\right)\cdot \phi\left( \frac{\bar{s}_{i,r}}{\sqrt{1-|\rho_{i,-r}|^2}} \right),
\label{KeyIdentity}
\end{equation}
where $\bar{s}_{i,r}=\bar{s}_i - \rho_{i,-r}'\xx_{-r}$.
Now, it can be checked that we can rewrite
$$\rho_i'\xx - |\rho_i|^2\bar{s}_i = \sum_{r=1}^p \left[\rho_{ir}x_r (1-|\rho_{i,-r}|^2) - c_{ir}\right] ,$$
where we have set
$$ c_{ir}:= \rho_{ir}^2\bar{s}_i - \rho_{ir}^2 \rho_{i,-r}' \xx_{-r}.$$
We deduce $ d\pi_1 = \sum_{r=1}^p d\pi_{1,r},$ with
$$ \frac{d\pi_{1,r}}{dt} = \sum_{\bdelta(T)} \psi(\bdelta (T) ) \sum_{i=1}^n \int \Prod_{j\neq i}
 p_{j|\xx}^{\delta_j}q_{j|\xx}^{1-\delta_j}.(2 \delta_i-1)
 \frac{\sigma_i^2 }{2}\cdot
 \frac{\phi(d_i)[\rho_{ir}x_r(1-|\rho_{i,-r}|^2) - c_{ir}]}{\phi(\bar{s}_i)^2 (1-|\rho_i|^2)^{3/2}}  \prod_{k=1}^p\phi(x_k) \, dx_k .$$
But, by integrating w.r.t. the variable $x_r$ and by invoking equation~(\ref{KeyIdentity}), we observe that
\begin{eqnarray*}
\lefteqn{  \int_{-\infty}^{x_r} \phi(d_i)[\rho_{ir}u_r(1-|\rho_{i,-r}|^2) - c_{ir}] \phi(x_r)\, du_r }\\
&=& - \rho_{ir} (1-|\rho_i|^2)
\phi\left(\sqrt{\frac{1-|\rho_{i,-r}|^2}{1-|\rho_{i}|^2}} \cdot
\{ x_r - \frac{\rho_{i,r}\bar{s}_{i,r}}{1-|\rho_{i,-r}|^2} \}\right)\cdot \phi\left( \frac{\bar{s}_{i,r}}{\sqrt{1-|\rho_{i,-r}|^2}} \right) \\ &=& - \rho_{ir} (1-|\rho_i|^2) \phi(d_i)\phi(x_r).
\end{eqnarray*}
Thus, by an integration by parts w.r.t. $x_r$, we get
\begin{eqnarray*}
\lefteqn{ \frac{d\pi_{1,r}}{dt} = -\sum_{\bdelta(T)} \psi(\bdelta (T) ) \sum_{i=1}^n \sum_{j \neq i} \int \Prod_{k\neq i,j}
 p_{k|\xx}^{\delta_j}q_{k|\xx}^{1-\delta_j}.(2 \delta_i-1)(2 \delta_j-1)
 \frac{\sigma_i^2 }{2}\cdot \frac{\partial p_{j|\xx}}{\partial x_r} }\\
 &\cdot &
 \frac{\rho_{ir} (1-|\rho_i|^2) \phi(d_i)}{\phi(\bar{s}_i)^2 (1-|\rho_i|^2)^{3/2}}  \prod_{k=1}^p\phi(x_k) \, dx_k  \\
 &=& -\sum_{\bdelta(T)} \psi(\bdelta (T) ) \sum_{i\neq j} \int \Prod_{k\neq i,j}
 p_{k|\xx}^{\delta_j}q_{k|\xx}^{1-\delta_j}.(2 \delta_i-1)(2 \delta_j-1)
 \frac{\sigma_i^2}{2} \\
 &\cdot &
 \frac{\rho_{ir}\rho_{jr} \phi(d_i)\phi(d_j)}{\phi(\bar{s}_i)^2 (1-|\rho_i|^2)^{1/2}(1-|\rho_j|^2)^{1/2}}  \prod_{k=1}^p\phi(x_k) \, dx_k .
\end{eqnarray*}
And then, by summing up on $r$, we obtain
\begin{eqnarray}
\lefteqn{ \frac{d\pi_1}{dt}
= -\sum_{\bdelta(T)} \psi(\bdelta (T) ) \sum_{i< j} \int \Prod_{k\neq i,j}
 p_{k|\xx}^{\delta_j}q_{k|\xx}^{1-\delta_j} \nonumber}\\
 &\cdot &
 \frac{(2 \delta_i-1)(2 \delta_j-1)\phi(d_i)\phi(d_j)}{2(1-|\rho_i|^2)^{1/2}(1-|\rho_j|^2)^{1/2}}
 \rho_{i}'\rho_{j} [\frac{\sigma_i^2}{\phi(\bar{s}_i)^2} + \frac{\sigma_j^2}{\phi(\bar{s}_j)^2}]  \prod_{k=1}^p\phi(x_k) \, dx_k .
\hspace{3cm}
\label{Gendpi1}
\end{eqnarray}
By summing the equations~(\ref{Gendpi1}) and~(\ref{Gendpi2}) up, we get the result. $\Box$

\section{Proof of theorem~\ref{FirstpTD}.}
\label{DemFPTD}
Let $ \psi_p(\bdelta) = \min(\sum_{i=1}^n \delta_i,p).(1-R),$
and $ \theta_p(\bdelta) = \1(\sum_{i=1}^n \delta_i \leq p).(1-R)$.
We can checked easily that, for every $p\geq 1$,
$$ \psi_p = \psi_{p-1} + 1 -R - \theta_{p-1}.$$
The drift of the value process associated with
a particular payoff $\psi$ will be denoted by $d\pi(\psi)$.
Obviously, $\pi$ is a linear functional, and $d\pi(\psi)=0$ if the payoff is risk-free. We deduce
\begin{equation}
d\pi(\psi_p) = d\pi(\psi_{p-1}) - d\pi(\theta_{p-1}).
\end{equation}
Now, let us specify $d\pi(\theta_p)$ for $p\leq n$.
To lighten the notations, we set
$$ \tilde A_{ij}= \frac{(1-R)}{2}\left[2 \beta_{it} \beta_{jt} \rho_{ij}  -\rho_i\rho_j \beta_{it}^2  - \rho_i \rho_j \beta_{jt}^2 \right]
\frac{  \phi(d_i)\phi(d_j)\phi(x)}{\sqrt{1-\rho_i^2}\sqrt{1-\rho_j^2}}\cdot$$
Let us apply theorem~\ref{theocopG}.
By distinguishing between the different values that are taken by the couples $(\delta_i,\delta_j)$, we get
\begin{eqnarray*}
\lefteqn{ d\pi_t(\theta_p) =
\sum_{\bdelta(T)} \1(\sum_{l=1}^n \delta_l\leq p) \sum_{i<j} (2\delta_i-1).(2\delta_j-1)\int \Prod_{k\neq i,j}
 p_{k|x}^{\delta_k}q_{k|x}^{1-\delta_k} \tilde A_{ij} \, dx \, dt }\\
&=& \sum_{i<j} \sum_{\bdelta_{-(i,j)}} \left[ \1(\sum_{l\neq i,j} \delta_l\leq p-2)
- 2 \1(\sum_{l\neq i,j} \delta_l\leq p-1) + \1(\sum_{l\neq i,j} \delta_l\leq p) \right]
\cdot \int \Prod_{k\neq i,j}  p_{k|x}^{\delta_k}q_{k|x}^{1-\delta_k} \tilde A_{ij} \, dx \, dt ,
\end{eqnarray*}
where $\bdelta_{-(i,j)}$ denotes the vector of $(\delta_k)$, with $k\in \{1,\ldots,n\}$ but $k \neq i,j$.
To simplify, let us denote
$$ \alpha_p := \sum_{i<j} \sum_{\bdelta_{-(i,j)}} \1(\sum_{l\neq i,j} \delta_l= p)  \int \Prod_{k\neq i,j}
 p_{k|x}^{\delta_k}q_{k|x}^{1-\delta_k} \tilde A_{ij} \, dx \, dt.$$
If $p\geq 1$, we have got
\begin{align*}
d\pi_t(\theta_p) &=  \sum_{i<j} \sum_{\bdelta_{-(i,j)}} [ \1(\sum_{l\neq i,j} \delta_l =  p)   -\1(\sum_{l\neq i,j} \delta_l = p-1)]
\int \Prod_{k\neq i,j}  p_{k|x}^{\delta_k}q_{k|x}^{1-\delta_k} \tilde A_{ij} \, dx \, dt , \\
&= \alpha_p -\alpha_{p-1}.
\end{align*}
We deduce, for all $p\geq 2$,
$$  d\pi(\psi_p) = d\pi(\psi_{p-1}) - \alpha_{p-1} +\alpha_{p-2}.$$
By iterating the latter equation, we get $  d\pi(\psi_p) = d\pi(\psi_{1}) - \alpha_{p-1} +\alpha_{0}$.
But $\pi(\psi_1)$ is the hedging error of the First To Default, that is equal to $-\alpha_0$
(see equation~(\ref{FTDpi})).
Finally, we get $ d\pi(\psi_p) = - \alpha_{p-1},$ so the result. $\Box$

\section{Proof of theorem~\ref{MartingaleCondProb}}
\label{ProofMarting}
In the one-period Merton model, with our notations and given $\Hc_t$,
\begin{align*}
p_{j,t|x}&=Q\left( \tau_j \leq T | \Hc_t \right)= Q\left( \int_0^T \psi_j(u) \, dW_{ju} < \bar{b}_{jT} |  \Hc_t \right)   \\
&= Q\left( \sqrt{1-\rho_j^2}\int_t^T \psi_j(u) \, dW^\ast_{ju} < \bar{b}_{jT} - \int_0^t \psi_j(u)\, dW_{ju} - \rho_j \int_t^T \psi_j(u)\, dW_{u} |  \Hc_t \right) \\
&= \Phi \left( \frac{\bar{b}_{jT} - \int_0^t \psi_j(u)\, dW_{ju} - \rho_j \int_t^T \psi_j(u)\, dW_{u}}{\sqrt{1-\rho_j^2}\sqrt{T-t}} \right):=\Phi(d_{jt}^*).
\end{align*}
When $\psi_j(u)=1$ for any $j$ and $u$, we recover the standard
Merton model and the pricing formula~(\ref{GCMSpecif}). In this
case, $\Hc_t=\sigma(W_{ju},W_T-W_u,u\leq t)$. For any $t,t'$, $T\geq
t>t' \geq 0$, we get

\begin{eqnarray*}
\lefteqn{ E[p_{j,t|x} |\Hc_{t'}] = E\left[ \Phi \left( \frac{\bar{b}_{jT} - W_{jt} - \rho_j (W_{T} - W_t)}{\sqrt{1-\rho_j^2}\sqrt{T-t}} \right) |\Hc_{t'} \right] }\\
&=& E\left[ \1\left( \sqrt{1-\rho_j^2}\sqrt{T-t}Z + (W_{jt} - W_{jt'}) - \rho_j ( W_{t} - W_{t'}) \leq  \bar{b}_{jT} - W_{jt'} - \rho_j ( W_{T} - W_{t'}) \right) |\Hc_{t'} \right] \\
&=& E\left[ \1\left( \sqrt{1-\rho_j^2}\sqrt{T-t}Z + \sqrt{1-\rho_j^2}(W^*_{jt} - W^*_{jt'}) \leq  \bar{b}_{jT} - W_{jt'} - \rho_j ( W_{T} - W_{t'}) \right) |\Hc_{t'} \right],
\end{eqnarray*}
for some gaussian r.v. $Z$, independent of $(W_{jt})$ and $(W_t)$.
We get
$$E[p_{j,t|x}|\Hc_{t'}] = \Phi \left( \frac{\bar{b}_{jT} - W_{it'} - \rho_j ( W_{T} - W_{t'})}{\sqrt{1-\rho_j^2}\sqrt{T-t'}} \right)= p_{j,t'|x}.$$
Therefore, the process $(p_{jt|x=W_T - W_t})_{t\in [0,T]}$ is an $\Hc$-martingale under $Q$. $\Box$


\section{Proof of theorem~\ref{Archim}}
In an Archimedean framework, we would like to find some dynamics that allow a cancellation of the hedging error $d\pi_t$ over all trajectories
$(Q_{it})$, $i=1,\ldots,n$, $t\in [0,T]$ and for all payoff $\psi$. It implies we have to satisfy
equation~(\ref{GeneralCondition}) by the same arguments as in the proof of theorem~\ref{UniqueDynamics}.
Particularly, we assume the case $Q_{it}=Q_{jt}$ for all $t$ and for some couple $(i,j)$, $i\neq j$~\footnote{Even if this event occurs with probability zero,
the latter assumption can be justified. It is sufficient to consider trajectories s.t. $\| Q_{it} - Q_{jt} \|<\varepsilon$ for some arbitrarily small $\varepsilon > 0$.}
Since the volatilities are the same measurable functions of their trajectories $(Q_{it})$,
this implies that $\sigma_{it} = \sigma_{jt}$ on these particular trajectories.
Thus, we have to satisfy the necessary relation
\begin{equation}
a \left(\frac{\partial p_{i|x}}{\partial Q_i}\right)^2 f_X(x) = \chi_i(x) \frac{\partial p_{i|x}}{\partial x}
\end{equation}
for some non zero constant $a$.
By simple calculations, it is equivalent to
$$ a x^2 p_{i|x} f_X(x) = \left[\xi'(\bar \xi_i) \right]^2 \bar \xi_i \chi_i(x) ,$$
for all $x$ in the support of $X$ and every $Q_i \in [0,1]$.
By deriving w.r.t $x$ and by making the change of variable $y=\bar\xi_i$, we have to satisfy
\begin{equation}
 a [ 2 - y x + x\frac{f_X'}{f_X}(x) ]  = y[x+ \frac{\xi''}{\xi'}(y)] ,
\label{condArch}
\end{equation}
for all $x$ in the support of $f_X$ and for all $y\in [0,1]$.
Particularly, when $y$ tends to zero, $\frac{\xi''}{\xi'}(y)$ tends to $-E[X^2]/E[X]$ that is finite by assumption. Thus, we have to
satisfy
$$ x\frac{f_X'}{f_X}(x) + 2 =0,$$
for all $x\geq 0$.
This implies the support of the density function $f_X$ contains some interval $[\alpha,\beta]\subset \RR^+$, where it is equal to
$$ f_X(x)= \frac{\alpha }{x^2}, $$
for some positive constant $\alpha$.
But then, equation~(\ref{condArch}) implies $a=(-1)$ and $\xi''=0$, that cannot be satisfied clearly. $\Box$

\section{Proof of theorem~\ref{Clayton}}
\label{DemClayton}

Under our assumptions, we calculate
$$\frac{\partial p_{i|x}}{\partial Q_i}= -x\theta p_{i|x} (1-Q_{it})^{-\theta-1}      ,$$
$$\frac{\partial^2 p_{i|x}}{\partial^2 Q_i}=(x\theta)^2 p_{i|x} (1-Q_{it})^{-2\theta-2}
-x\theta (\theta+1) p_{i|x} (1-Q_{it})^{-\theta-2},$$
$$\frac{\partial p_{i|x}}{\partial x}=  p_{i|x} [1- (1-Q_{it})^{-\theta}].$$
By an integration by part, we obtain the identity
$$ \chi_i(x)=  \frac{\partial p_{i|x}}{\partial Q_i} \cdot \frac{x\theta f_X(x)}{1-Q_{it}} \cdot $$
Therefore, with our choices of volatilities $\sigma_{it}$ and the underlying processes,
it can be checked easily we satisfy~(\ref{GeneralCondition}), replacing $\rho_{ij}$ by
$$<dQ_{it},dQ_{jt}>=\sigma_{it}\sigma_{jt} \beta_i' (Q_{it})\beta_j(Q_{jt}).$$
Finally, we do not have to take into account the so-called term $\eta$ of the theorem~\ref{thdpit2}, because $\lim_{x\rightarrow 0} \chi_i(x)=0$ and $\lim_{x\rightarrow +\infty} \chi_i(x)=0$.
$\Box$

\section{Proof of theorem~\ref{GCMTheTrue}}
\label{DemGCMTheTrue} To lighten the notations, we set
$\eta_i:=\eta\left(\alpha_i x - h(Q_{it}) \right)$, $h_i:=h(Q_{it})$
and $f:=f_X(x)$. Under our assumptions, we get easily
$$ \frac{\partial p_{i|x}}{\partial x} = \eta_i' \alpha_i,\;\;
\frac{\partial p_{i|x}}{\partial Q_i} :=-\eta_i' h'_i,\;\;
\frac{\partial^2 p_{i|x}}{\partial^2 Q_i} :=\eta_i'' (h'_i)^2 - \eta_i' h''_i.$$
We would like to satisfy the partial differential
equation~(\ref{GeneralCondition}) that is rewritten here
\begin{equation}
2(\sigma_i h'_i).(\sigma_j h'_j) \rho_{ij} f = \alpha_j\sigma_i^2\frac{\chi_i}{\eta'_i}
+\alpha_i\sigma_j^2\frac{\chi_j}{\eta'_j}\cdot
\label{EDPGCMTheTrue}
\end{equation}
The latter equation has to be satisfied for all the triplets
$(Q_i,Q_j,x)$. Actually, it is a very strong constraint, because it
can be proved that this implies: there exist some constants $C_i$
and $\bar C(i,j)$, $i,j=1,\ldots,n$, such that
\begin{eqnarray}
\sigma_i h'_i & =&  C_i, \label{MyEq1}\\
\sigma_i^2\frac{\chi_i}{\psi'_i}\alpha_j & =  & \bar C(i,j) \rho_{ij}f,  \label{MyEq2}
\end{eqnarray}
for all indices $(i,j)$, $i\neq j$. The latter result is a
consequence of the following lemma:
\begin{lemma}
Consider some real functions $a,b,c,d$ such that
$$ a(x)b(y)=c(x)+d(y),$$
for all couples $(x,y)$. Then these four functions are constant.
\label{MyLemma}
\end{lemma}
This lemma is proved at the end of the current proof. Note that
$C_i$ is not a function of $x$ because $\sigma h'_i$ is a function
of $Q_i$ only (not of $x$). Moreover, equation~(\ref{MyEq2}) is due
to the fact we can consider the particular case $\alpha_i=\alpha_j$
and some trajectories where $Q_{it}=Q_{jt}$~\footnote{In this case,
the two members of the rhs of equation~(\ref{MyEq1}) are equal.}.

\medskip

From the latter equations, we get the existence of a constant
$\mu=\mu(i,j)$ such that
$$ \chi_i=\mu f \eta'_i (h'_i)^2.$$
By differentiating this equation w.r.t. $x$, we get
\begin{equation}
 \frac{\eta''_i}{\eta'_i}-\frac{h''_i}{(h'_i)^2} = \mu \left[\frac{f'}{f}+\alpha_i\frac{\eta''_i}{\eta'_i}   \right].
\label{eta0}
\end{equation}
Now, by integrating the latter equation w.r.t. $x$, we obtain
\begin{equation}
 \eta'_i = \exp\left(\nu_i + \frac{x h''_i}{(1-\mu\alpha_i)(h'_i)^2}\right)f^{\mu/(1-\mu\alpha_i)},
\label{eta1}
\end{equation}
for some function $\nu_i:=\nu(Q_i)$. But~(\ref{eta0}) can be
multiplied by $h'_i$. And after an integration w.r.t. $Q_i$, we get
\begin{equation}
 \eta'_i = \exp\left(\xi(x) - \frac{[\ln(h'_i)+\mu_i f' h_i/f]}{(1-\mu\alpha_i)}\right),
\label{eta2}
\end{equation}
for some function $\xi$. If $f'/f$ were a constant function, the
density $f$ would be exponentially distributed and the support of
$X$ cannot be $\RR$ as a whole. Moreover, $h$ is a not a constant by
assumption. Thus, equations~(\ref{eta1}) and~(\ref{eta2}) together
imply that the crossed terms $f'h_i/f$ and $xh''_i (h'_i)^{-2}$ are
proportional. In other words, this means that $f'/f$ (resp. $h_i
h'_i$) is proportional to $x$ (resp. $h''_i /h'_i$). Therefore, by
integration, we check that the usual Gaussian copula model is the
single possible model that satisfies such requirements. $\Box$

\medskip

To finalize this proof, it remains to state lemma~\ref{MyLemma}.
This is done now.

\medskip

If there exists some real number $x$ such that $a(x)=0$, then it is
clear that $c$ and $d$ are constant functions, and then $a$ and $b$
too. We can lead the same reasoning with $b$. Thus, we can assume
that $a$ and $b$ cannot be zero. Now, assume that $b$ is not a
constant: there exist a couple $(y_1,y_2)$ such that $b(y_1)\neq
b(y_2)\neq 0$. Then, for every real number $x$, we have
$$ \frac{b(y_1)}{b(y_2)} = \frac{c(x)+d(y_1)}{c(x)+d(y_2)},$$
that is constant function in $x$. This implies that $c$ is a
constant function of $x$, and that $d=\alpha b+\beta$ for some
constants $\alpha$ and $\beta$. We deduce that $a(x)=(c+\alpha
b(y)+\beta)/b(y)$, so $a$ and $b$ are constant functions.

This is a contradiction with our initial assumption. Since this
reasoning can be lead with $a$ instead of $b$ similarly, we have
proved our lemma. $\Box$

\pagebreak

\oddsidemargin=0pt

\end{document}